\documentclass[fleqn,usenatbib]{mnras}

\usepackage{newtxtext,newtxmath}

\usepackage[T1]{fontenc}

\DeclareRobustCommand{\VAN}[3]{#2}
\let\VANthebibliography\thebibliography
\def\thebibliography{\DeclareRobustCommand{\VAN}[3]{##3}\VANthebibliography}

\usepackage{graphicx}	
\usepackage{amsmath}	

\newcommand{\msun}{M$_{\sun}$}
\newcommand{\mzams}{M$_\mathrm{ZAMS}$}

\title[The Pre-MS in SPS Models]{What do we mean by stellar mass? The impact of the pre-main sequence on the mass to light ratio of young and intermediate age stellar populations}

\author[Stanway et al.]{
Elizabeth R. Stanway, Conor M. Byrne and Ankur Upadhyaya 
\\
Department of Physics, University of Warwick, Gibbet Hill Road, Coventry CV4 7AL, UK\\
}

\date{Accepted 2025 August 21. Received 2025 August 19; in original form 2025 July 17}

\pubyear{2025}

\begin{document}
\label{firstpage}
\pagerange{\pageref{firstpage}--\pageref{lastpage}}
\maketitle

\begin{abstract}
  Stellar population synthesis models are an essential tool with which galaxy physical parameters are extracted from observations. However they are built on assumptions designed for use in the local Universe, and not always appropriate to high redshift galaxies. Here we consider the impact of including the hitherto-neglected stellar pre-main sequence delay timescale on the interpretation of composite stellar populations at ages of $<1$\,Gyr. We find that doing so has an impact on the optical luminosity of very young stellar populations of up to $\sim10$ per cent, although smaller changes in observed light ($<5$ per cent) are expected in most use cases. However the impact on the inferred stellar mass and mass-to-light ratios is significant (a factor of 2 or more), depending on how those properties are defined. We find that the short time scales for star formation in the distant Universe require a clearer definition for the stellar mass in a population, and will impact assumptions about the inferred shape of the stellar initial mass function from observations.
    
\end{abstract}




\section{Introduction}

Stellar Population Synthesis (SPS) models are foundational to our understanding of galaxy evolution. By modelling the spectral evolution of a population of stars, they permit synthetic observations to be compared to unresolved galaxies, and thus form a pathway towards interpreting light in terms of the physical properties of the galaxy generating it. SPS models provide the backbone for procedures such as galaxy spectral energy distribution (SED) fitting \citep[see e.g.][]{2012MNRAS.422.3285P,2013ARA&A..51..393C,2025MNRAS.540.2703B}, and the calibration for simpler prescriptions such as star formation rate indicators and galaxy mass-to-light ratios \citep[e.g.][]{2012ARA&A..50..531K,1979ARA&A..17..135F}. SPS models are thus used across cosmic time to translate an observed luminosity function into two more fundamental relations - the cosmic star formation rate density evolution and the growth of the stellar mass density in the Universe \citep[see e.g.][]{2014ARA&A..52..415M}.  As a result, they also provide the calibration for key observational constraints on the large simulations often used to understand the processes by which large galaxies like our own evolve. 

SPS models have existed in some form since the late 1960s \citep{1968ApJ...151..547T,1971Ap&SS..12..394T,1973ApJ...186...35T, 1976ApJ...203...52T, 1983ApJ...273..105B,1998MNRAS.300..872M}. At their simplest, they consider a population of coeval stars formed in a single star formation event and thus populating an initial mass function (IMF). As the population ages, the evolution of its contents is tracked, using either stellar evolution tracks or the isochrones derived from them. In spectral synthesis models \citep[e.g.][]{2005MNRAS.362..799M,2011MNRAS.418.2785M,1999ApJS..123....3L, 2011AJ....141...37L, 1993ApJ...405..538B, 2003MNRAS.344.1000B, 2009MNRAS.400.1019E,2012MNRAS.419..479E,2017PASA...34...58E}, the resulting simple stellar population (SSP) is combined with a grid of spectra from a stellar spectral library to create a synthetic spectrum which represents the integrated light emitted by the unresolved stellar population.

In the era of computational astrophysics SPS models have come of age, with improvements in the number of stars being modelled and the details and interior physics of those models \citep[e.g.][]{2017PASA...34...58E,2023ApJS..264...45F}. Extensive work has also explored the impact on synthetic stellar populations of prescriptions used for the stellar distribution such as initial mass distribution (IMF), binary population parameters (such as initial period or mass ratio distributions) or stellar metallicity \citep[e.g.][]{2019A&A...621A.105S,2020MNRAS.495.4605S,2020MNRAS.497.2201S,2023MNRAS.522.4430S,2021ApJ...922..177M,2022MNRAS.514.5706J,2022MNRAS.512.5329B,2023MNRAS.521.4995B}. Studies have also explored the uncertainties arising when SPS models are combined with star formation and metallicity histories to simulate the stellar populations of entire galaxies, and when those populations are paired with spectra from a stellar spectral library
to generate synthetic spectra for the population as a whole \citep[e.g.][]{2024arXiv241017697R,2025MNRAS.540.2703B,2020MNRAS.495..905R,2023MNRAS.525.5720J, 2023ApJ...944..141P,2019ApJ...873...44C,2019ApJ...876....3L}.

Such studies have shown that one of the most robustly recovered parameters of a galaxy based on optical or near-infrared photometry is its stellar mass since, for stellar populations exceeding around 100 Myr in age, this scales almost linearly with optical/near-IR luminosity in most SPS models. The impact of considerations such as the galaxy star formation history or the stellar spectra adopted for comparison are typically small effects, causing uncertainties of order 0.1 dex on the derived stellar mass \citep[e.g.][]{2022MNRAS.514.5706J} 

However there are known uncertainties even in stellar masses derived from mass to light ratios or SED fitting. Systematic offsets between masses reported for the same object when fitting with different SPS models arise primarily due to different shapes adopted for the low-mass stellar IMF (which dominates the mass but not the optical light in most stellar populations). Where the luminosity of SPS models are scaled to a birth mass, a correction must also be made for the stellar mass initially assigned to massive stars which may no longer be present in the population at late ages, and for the mass of the population of low temperature sub-stellar brown dwarfs ($M<0.08$\,\msun) which may only be detectable in the thermal infrared. In most SPS models, and the SED fitting based on them, the  stellar mass derived is that of stars with ongoing energy generation from nuclear fusion, and excludes any inferred compact remnant and brown dwarf populations.

However, as we demonstrate in this paper, for young stellar populations an additional correction may be required, which, in turn, requires the definition of \textit{stellar mass} to be carefully considered.

SPS models have traditionally produced co-eval simple stellar populations, in which each star is understood to have begun its nuclear H-burning main sequence lifetime (i.e. reached the Zero Age Main Sequence, or ZAMS) at the same epoch of star formation. The age of the stellar population is calculated from this epoch. However this foundational assumption neglects an important and long-recognised aspect of stellar evolution. Both analytic calculations and simulations indicate that a star formation event occurs when a giant molecular cloud collapses and fragments creating protostellar cores \citep{1987ARA&A..25...23S}. These grow through competitive accretion and their mass distribution is modified by dynamical processes such as mergers or ejection from the region. Each core must reach a sufficient size and central density before ignition occurs, with a timescale that is strongly dependent on its ZAMS mass \citep[see e.g.][]{1965ApJ...141..993I,1966ARA&A...4..171H,1995ApJ...446L..35B,1993ApJ...418..414P}. For stars above approximately 5\,\msun\ this is a rapid process, and ignition occurs within 1 Myr of the onset of collapse. However for lower mass stars it is a substantially extended process, with pre-main sequence (pre-MS) models for a 0.5\,\msun\ object extending beyond 100 Myr in pre-MS lifetime. 

In this paper we consider the impact of including a prescription for mass-dependent pre-main sequence delay in a population and spectral synthesis model which incorporates both single star and binary stellar evolution. In section \ref{sec:seps}, we combine pre-main sequence delays from Tognelli et al (2011) with stellar evolution tracks, population and spectral synthesis prescriptions from the Binary and Population Synthesis (BPASS) framework \citep{2009MNRAS.400.1019E,2012MNRAS.419..479E, 2017PASA...34...58E, 2018MNRAS.479...75S, 2022MNRAS.512.5329B}. Section \ref{sec:results} presents key results of the population and spectral synthesis, while section \ref{sec:discussion} considers their implication for observations of galaxies in both the distant and local Universe. Section \ref{sec:conclusions} summarises our conclusions and considers the challenging question of how we define stellar mass in galaxies.

Where necessary we adopt a flat $\Lambda$CDM cosmology with $H_0$=69.6\,km\,s$^{-1}$\,Mpc$^{-1}$, $\Omega_M=0.286$ and $\Omega_\Lambda=0.714$ \citep{2014ApJ...794..135B}.

\section{Stellar Evolution and Population Synthesis}\label{sec:seps}

\subsection{Pre-Main sequence evolution tracks}\label{sec:pmstracks}

\begin{figure}
    \centering
    \includegraphics[width=\columnwidth]{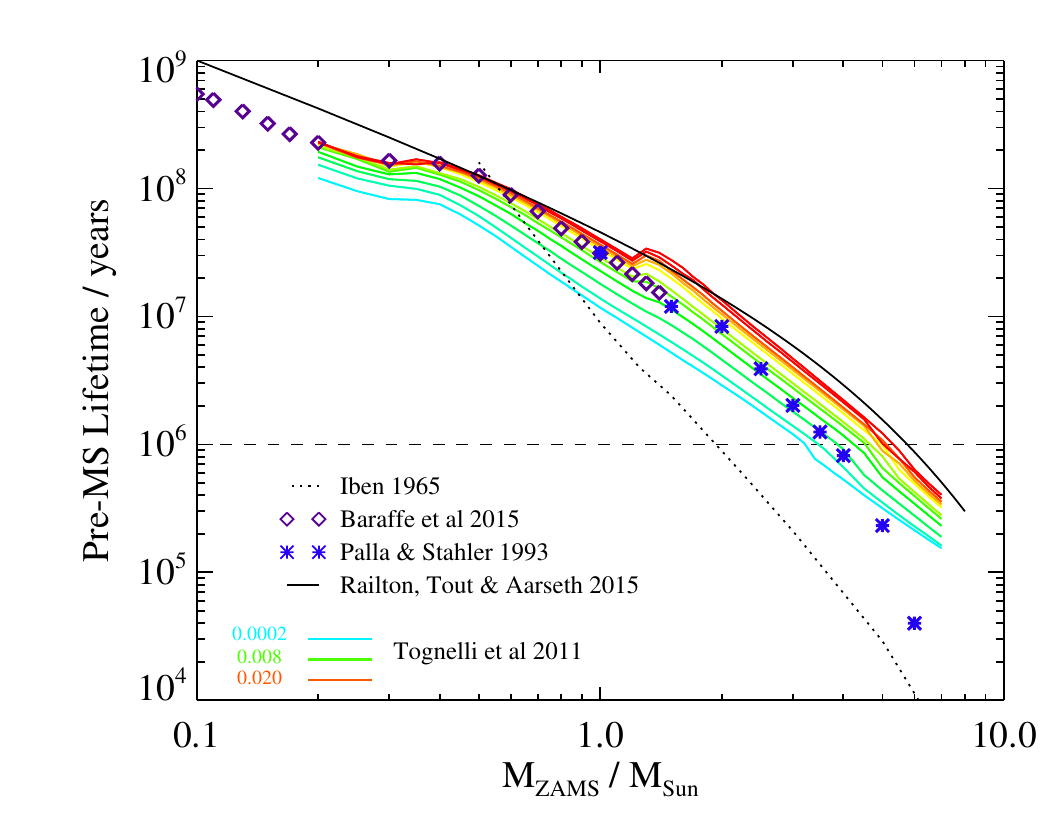}
    \caption{Pre-MS lifetime in a range of stellar models, as a function of \mzams. The horizontal line indicates the 1 Myr minimum age threshold adopted by many SPS models. The models of \citet{2011A&A...533A.109T} include a metallicity dependence for models between 0.01\,Z$_\odot$ (green) and 1.5\,Z$_\odot$ (red). The solid black line shows the pre-MS lifetime of models by \citet{2014PASA...31...17R}, the points those of \citet[diamonds]{2015A&A...577A..42B} and \citet[asterisks]{1993ApJ...418..414P} and the dotted black line the earlier work of \citet{1965ApJ...141..993I} for comparison.}
    \label{fig:prems}
\end{figure}

Models for the pre-main sequence of stars have converged over recent decades. While the pre-MS lifetimes of massive stars have been revised upwards since the early work of \citet{1965ApJ...141..993I} and \citet{1993ApJ...418..414P}, there is now generally reasonable agreement between independent model grids as demonstrated in Figure \ref{fig:prems}.

In the Figure, we show the time elapsed between the onset of star formation and the proto-star evolving onto the hydrogen-burning main sequence. For the stellar evolution tracks of \citet{2015A&A...577A..42B} and \citet{2011A&A...533A.109T}, which extend part way onto the hydrogen burning main sequence, we identify the pre-MS lifetime as the age at which a star first appears within 0.01\,dex of its main sequence temperature.  The \citet{2011A&A...533A.109T} tracks also demonstrate the variation with metallicity from low (blue) to high (red) values.

The Pisa pre-main sequence library was presented by \citet{2011A&A...533A.109T}. They calculated a grid of stellar evolution models with the FRANEC evolution code \citep{2008Ap&SS.316...25D,2009A&A...507.1541V}. We downloaded the model grid from the VizieR On-line Data Catalog J/A+A/533/A109. Models were calculated at a range of bulk metallicities from 0.01\,Z$_\odot$ to 1.5\,Z$_\odot$, for stars with ZAMS masses $M = 0.2-7.0$\,M$_\odot$), and were evolved through ages from 1 to 100 Myr. As Figure \ref{fig:prems} demonstrates, these pre-MS models show a strong variation of pre-MS lifetime on metallicity, such that low metallicity stars form more rapidly than their higher metallicity counterparts.

\subsection{BPASS stellar evolution tracks}\label{sec:bpasstracks}

The Binary Population and Spectral Synthesis (BPASS) framework is built on a custom grid of stellar evolution models calculated using the BPASS-STARS code \citep{2017PASA...34...58E, 2018MNRAS.479...75S}, an extensively modified version of the Cambridge STARS code \citep{1971MNRAS.151..351E,1995MNRAS.274..964P,2004MNRAS.353...87E}. Models are calculated at 13 metallicity mass fractions from Z=$1\times10^{-5}$ to Z=0.040, where Z$_\odot$=0.020. Single star evolution models are calculated in the zero-age Main Sequence mass range, \mzams=$0.1 - 300$\,\msun. All models are calculated with an elapsed age starting from the onset of hydrogen core burning. Binary models take primary stars in the same mass range, and secondaries with initial mass ratios in the range $q=0.1 - 0.9$ and initial periods in the range log(P/days)=$0 - 4$. After the primary star has reached supernova or collapsed to a white dwarf, a secondary model continues the evolution of any remaining binaries with a compact object remnant. We use the full grid of models calculated for BPASS v2.2 \citep{2018MNRAS.479...75S}.

\subsection{Population and Spectral Synthesis}\label{sec:bpass_sps}

The BPASS stellar tracks are combined given an initial mass function and binary parameter distribution, and the properties of the population are calculated at 42 age steps between 1\,Myr and 20\,Gyr from the onset of star formation. In all previous work the onset of hydrogen burning (ZAMS) of all stars in the population has been considered as co-eval. 

For the purposes of population and spectral synthesis, here we adopt the default BPASS v2.3.1 prescription of \citet{2023MNRAS.521.4995B}. 
We calculate results for a broken power law IMF, such that the number of stars scales as $dN/dM \propto M^\alpha$ with $\alpha=-1.3$ in the range $M=0.1-1$\,\msun\ and $\alpha=-2.35$ in the range $M=1-300$\,\msun. The integral of the initial mass function across this range is normalised to $10^6$\,\msun. Stars with masses below 0.1\,\msun, including binary companions, are not included in the mass budget.

\begin{figure*}
    \centering
    \includegraphics[width=0.98\columnwidth]{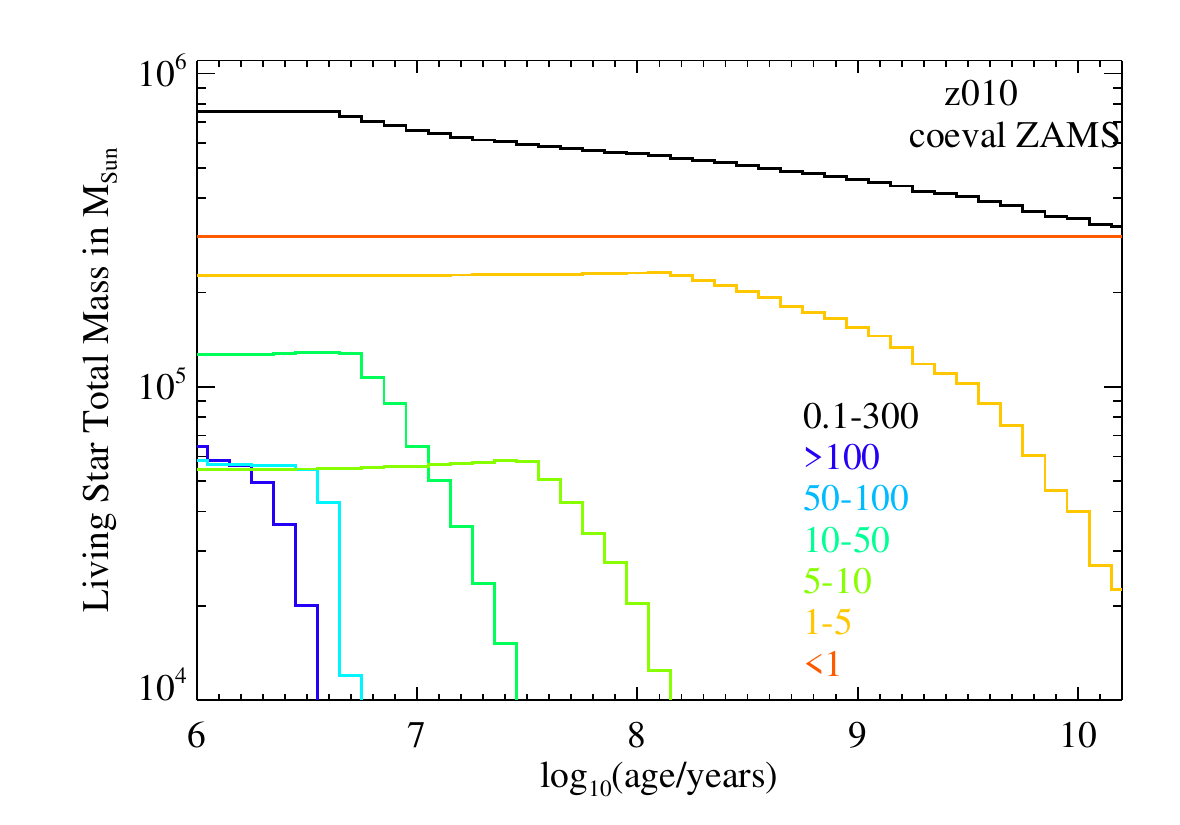}
    \includegraphics[width=0.98\columnwidth]{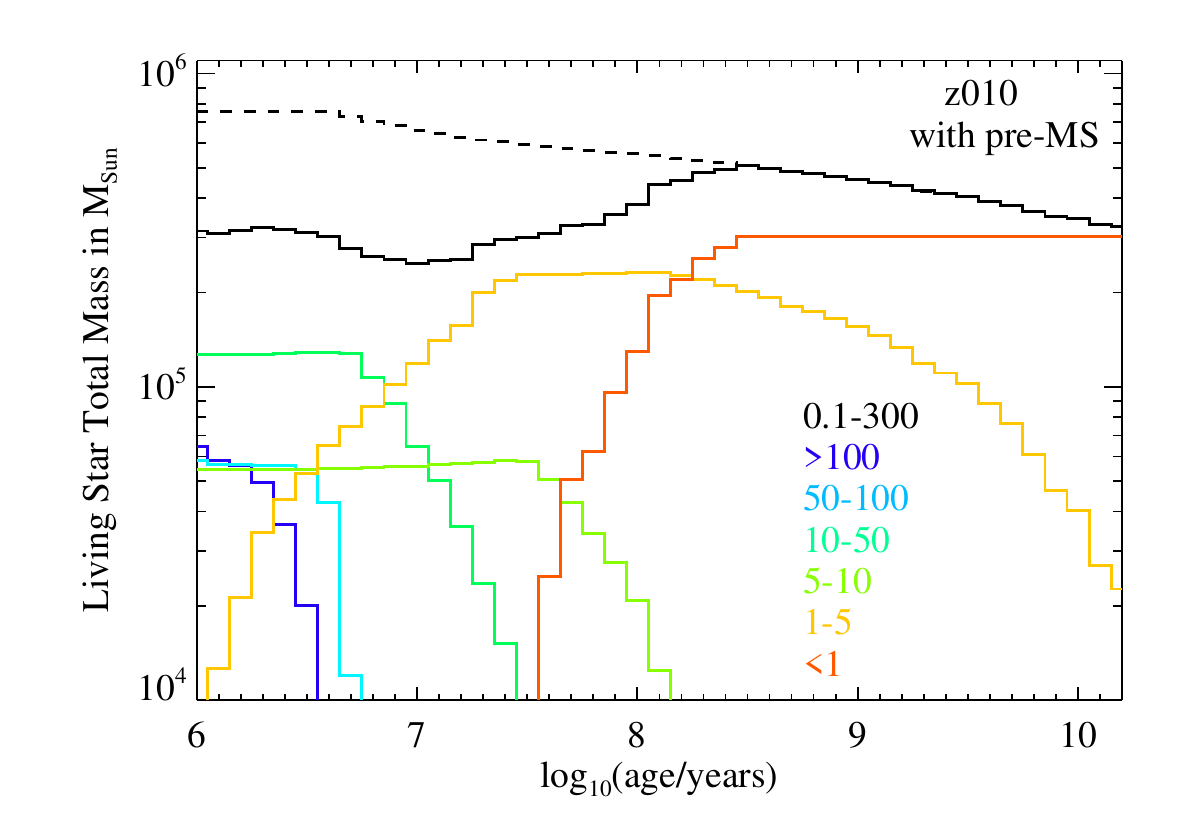}
    \caption{The energy-generating stellar mass as a function of elapsed time since the onset of star formation. The left hand panel and dashed line on the right shows the evolution of mass in a population in which all stars are assumed to share a common ZAMS epoch. The right hand panel shows the difference when a common onset of star formation is assumed together with pre-MS formation timescales. Each is further divided to show contribution by primary mass into ranges \mzams$<$1 (red), 1-5 (orange), 5-10 (light green), 10-50 (mid green), 50-100 (pale blue) and $>100$\,\msun (dark blue).}
    \label{fig:starmass}
\end{figure*}

For the purposes of spectral synthesis, the BPASS v2.3.1 prescription adopts spectra derived from the stellar atmosphere models of \citet{2016ApJ...823..102C} for stars on the main sequence and giant branch, the Potsdam Wolf Rayet (PoWR) atmosphere models for stripped and partially-stripped (e.g. WNH) stars \citep{2003A&A...410..993H,2015A&A...577A..13S}, the atmospheres of \citet{2017ApJS..231....1L} for white dwarfs and a small custom grid of models generated using WMBASIC for stars with $T>25$\,kK \citep{2017PASA...34...58E}. We use stellar atmospheres scaled to solar abundances (i.e. not $\alpha$-enhanced or otherwise varying in composition). Pre-MS stars, brown dwarfs below 0.1\,\msun, neutron stars and black holes are assumed to make no significant spectral contribution. We model a mixed single star and binary star population. For the latter, we use the binary fraction parameters of \citet{2017ApJS..230...15M}, implemented as described in \citet{2018MNRAS.479...75S}, independent of metallicity. We additionally correct the evolution of a small number of binary systems which were incorrectly flagged as mergers in earlier population synthesis versions, with negligible impact on the output populations.

To evaluate the impact of pre-main sequence delay times, we adopt the metallicity-dependent pre-MS lifetimes of \citet{2011A&A...533A.109T}, which were derived for metallicities in the range Z=0.0002-0.030 (0.01-1.5 Z$_\odot$) and masses M$_{ZAMS}$=0.2-7\,\msun. For BPASS stellar models which fall outside this range (at Z=$1\times10^{-4}$ and Z=0.040, or with M$_\mathrm{ZAMS}<$0.2\,\msun) we adopt the nearest available pre-MS grid point rather than extrapolating.  For models falling within the grid, we interpolate in both mass and metallicity. Models with a pre-MS lifetime shorter than log(age/years)=5.5, a third of the duration of our first model time bin, are considered to form with no delay, thus negating the need for extrapolation to higher masses. As Fig.~\ref{fig:prems} demonstrates, this leads to zero delay in our synthesis above a metallicity-dependent threshold which varies from 4\,\msun\ at low metallicities to 7\,\msun\ at near-Solar metallicities.

The effect of stellar companions on pre-MS lifetimes remains poorly understood. In the absence of other constraints, we make the assumption that the evolution of the initially more massive star (the primary) dominates over that of its companion, and assign the secondary star the same delay as its primary, regardless of mass ratio and initial separation.

One modification to the population synthesis prescription was required in this analysis. When a massive binary model reaches the end of the life of its primary, numerous different outcomes are possible as the result of supernova kicks, which occur with a range of velocities and in random directions. These can unbind the binary, harden its orbit or lead to an immediate merger between the remnant and the surviving secondary. In a default BPASS population synthesis, the impact of a large number of possible kicks \citep[typically 10,000 random samples drawn from the velocity distribution of][]{2005MNRAS.360..974H} is simulated. The resulting statistical distribution of outcomes is mapped onto a grid of pre-calculated secondary models which continue the evolution with a remnant, or as a single star. As a result many primary models (extending to the end of the first star) can map onto a single secondary model (and vice versa). 

Since here we require a known primary mass to set the pre-MS lifetime, we reduce the number of kick velocity samples to twenty for each supernova, and we record the pre-MS time delay associated with each primary star along with the weighting to be assigned to each of its selected secondary models. We note that this will only apply to massive stars (i.e. those undergoing core collapse at the end of nuclear burning) and their companions, for which the pre-MS delay time is typically smaller than the 1Myr `ZAMS' time step of BPASS population synthesis. To prevent for the finite grid of stellar models causing numerical artifacts as key stellar evolution phases (e.g. evolution onto the giant branch) cross arbitrary time bin boundaries, we show most time evolution results with a boxcar smoothing across three adjacent time bins.

\section{Results}\label{sec:results}

We calculate the integrated population parameters for two possible population syntheses: one in which the zero age main sequence of all stars is considered to be co-eval (as in previous work) and a second in which the onset of star formation is coeval, but each stellar model only begins contributing to the population after a mass-dependent time delay commensurate with its pre-main sequence lifetime.

We show most results for a metallicity of Z=0.010. Due to the metallicity dependence of stellar lifetimes and binary interactions, there are small differences in detail with metallicity, but no significant changes in the overall behaviour of the population, except where indicated. Both compact stellar remnants and pre-MS protostars are excluded from contributions to the population light and mass.

\subsection{Surviving stellar masses}\label{sec:res-mass}

In Figure \ref{fig:starmass} we show how the energy-generating total stellar mass evolves as a function of elapsed time since the onset of star formation for a simple stellar population. The left hand panel (and the dashed line on the right) shows the evolution of mass in a population in which all stars are assumed to share a common ZAMS epoch. The right hand panel shows the difference when a common onset of star formation is assumed together with pre-MS formation timescales. The distributions are further divided to show the relative contribution of stars by primary (or single star) mass in the ranges \mzams$<$1, 1-5, 5-10, 10-50, 50-100 and $>100$\,\msun. Secondary star masses are included with the mass contribution of their primary.  

In the canonical co-eval ZAMS scenario, stars with \mzams$<5$\,\msun\ dominate the stellar mass at all epochs, with the fractional contribution from massive and very massive stars declining rapidly with population age.  In the scenario where the onset of star formation is co-eval and pre-MS delay timescales are taken into account, the living stellar mass of young stellar populations (log(age/years)$<$7) is dominated by massive stars in the range \mzams=10-50\,\msun, with near equal contributions from very massive stars and stars in the range 1-10\,\msun.

\subsection{Mass to Light Ratios}\label{sec:mlratio}

A key property of an integrated stellar population is its mass-to-light ratio, whether this is used directly based on a single photometric data point, or implicitly through application of SED modelling and fitting algorithms.

\begin{figure}
    \centering
    \includegraphics[width=0.98\columnwidth]{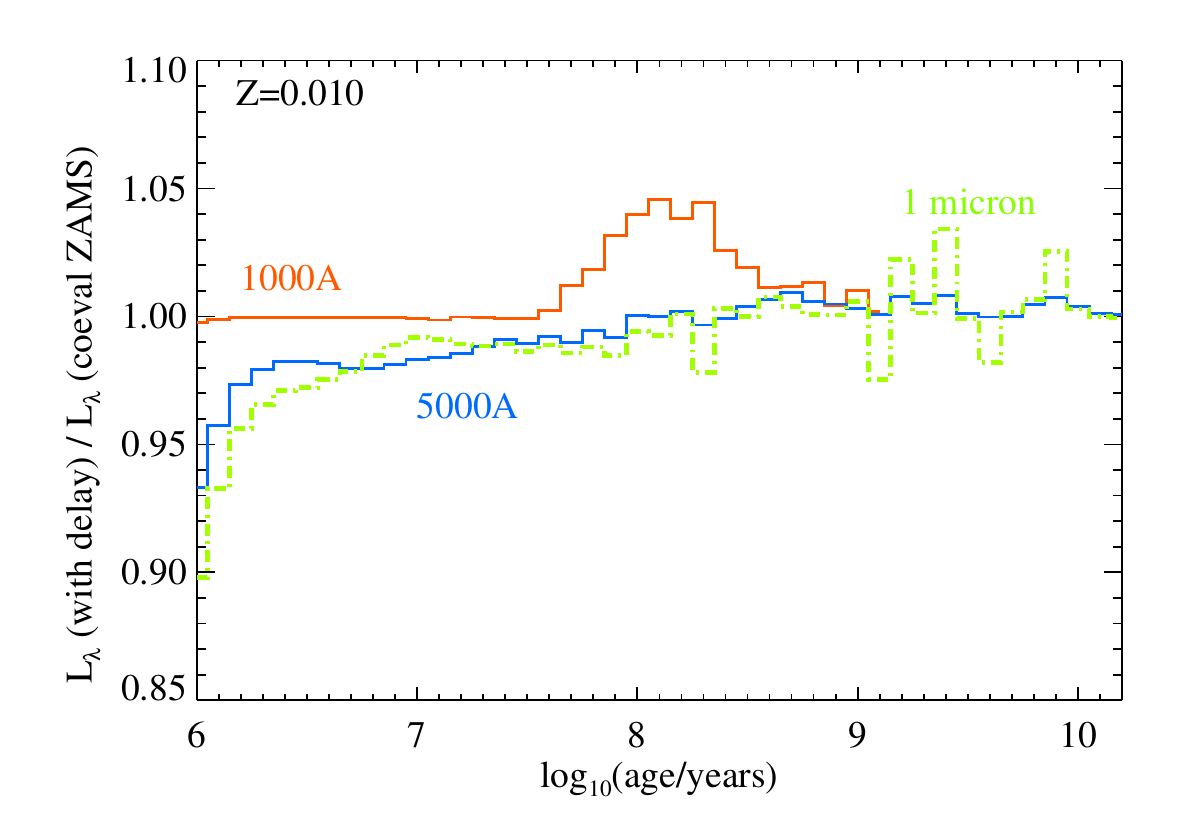}
   \includegraphics[width=0.98\columnwidth]{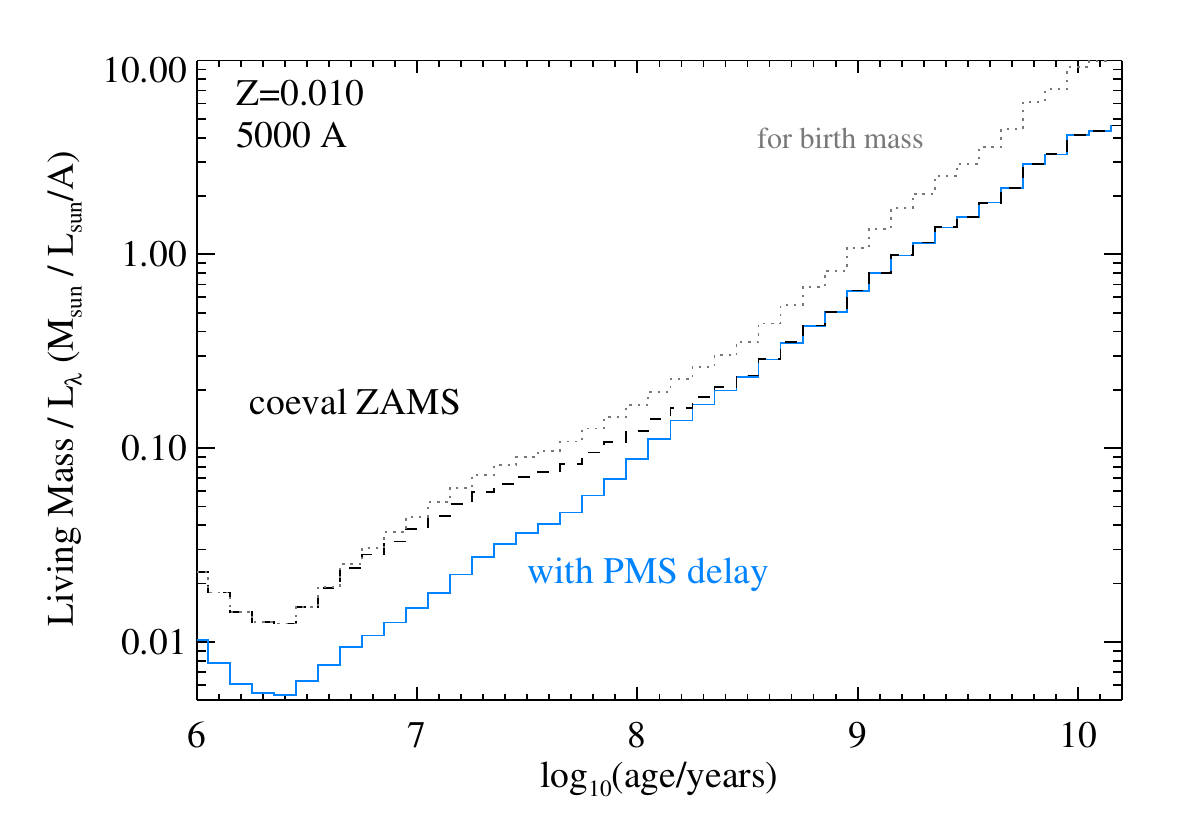}
    \includegraphics[width=0.98\columnwidth]{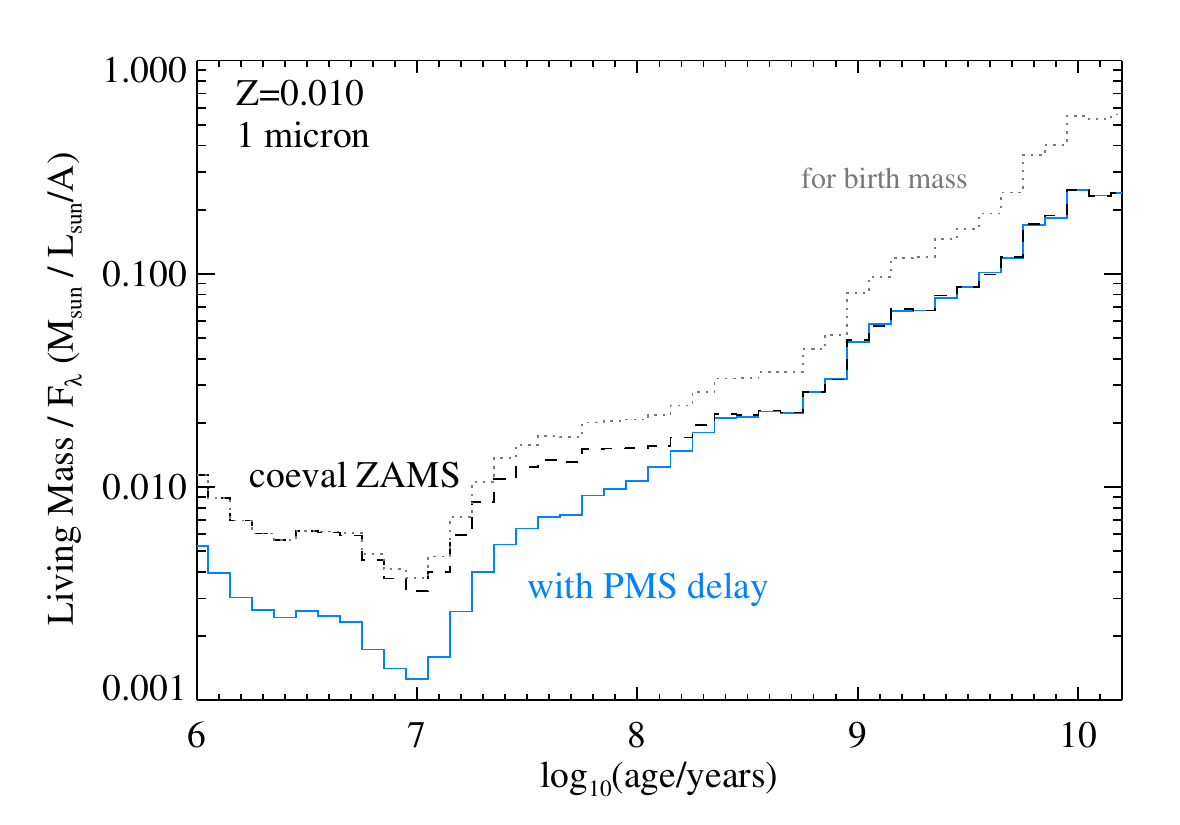}
    \caption{The effect of pre-main sequence delay timescales on the stellar mass to light ratio. The top panel shows the flux density ratio as a function of stellar population age in the ultraviolet, optical and infrared, comparing a coeval ZAMS scenario with models including a pre-main sequence delay. The lower two panels show the mass to light ratio given in solar masses per L$_\odot$/\AA\ calculated from L$_\lambda$ at 5000\,\AA\ and 1\,$\mu$m). The pale dotted line shows the equivalent if a constant stellar mass of $10^6$\,\msun\ (the total birth mass) is assumed.}
    \label{fig:mlratio}
\end{figure}

Fig.~\ref{fig:mlratio} demonstrates the impact of including a mass-dependent pre-MS delay phase on the true stellar mass-to-light ratio as a function of time in the optical (at 5000\,\AA) and infrared (at 1\,$\mu$m). Results have been smoothed across adjacent time bins to account for the finite grid of models in mass, binary mass ratio and binary period. As the upper panel demonstrates, at stellar population ages between 1.5\,and 300 Myr, the mean deviation in luminosity between the coeval ZAMS and Pre-MS delay scenarios is just 0.5 per cent in the optical and 0.75 per cent at 1 micron. In the far-ultraviolet (at 1000\,\AA) the deviation is effectively zero except at ages of $\approx$100\,Myr, at which point the total ultraviolet luminosity is very low and the later evolution of B and A stars has a slight effect.
However, despite the minor impact on luminosity, the strong dependence of the nuclear-burning stellar mass on scenario that was demonstrated in Fig.~\ref{fig:starmass} results in a difference in mass-to-light ratio  that varies from a factor of 2 at early times to effectively zero at an age of around 300\,Myr.

If, rather than the living stellar mass, the total mass of stars that \textit{will} be formed (i.e. $10^6$\,\msun\ in both cases) is used to calculate the mass-to-light ratio, this difference disappears, with the age evolution following the coeval ZAMS values at early times. However in this case, the rationale for omitting the mass lost from the nuclear-burning stellar population due to winds, supernovae and compact object formation is unclear. As a result, the stellar cluster formed-mass-to-light ratio will remain high at late times.

\begin{figure*}
    \centering
    \includegraphics[width=0.98\columnwidth]{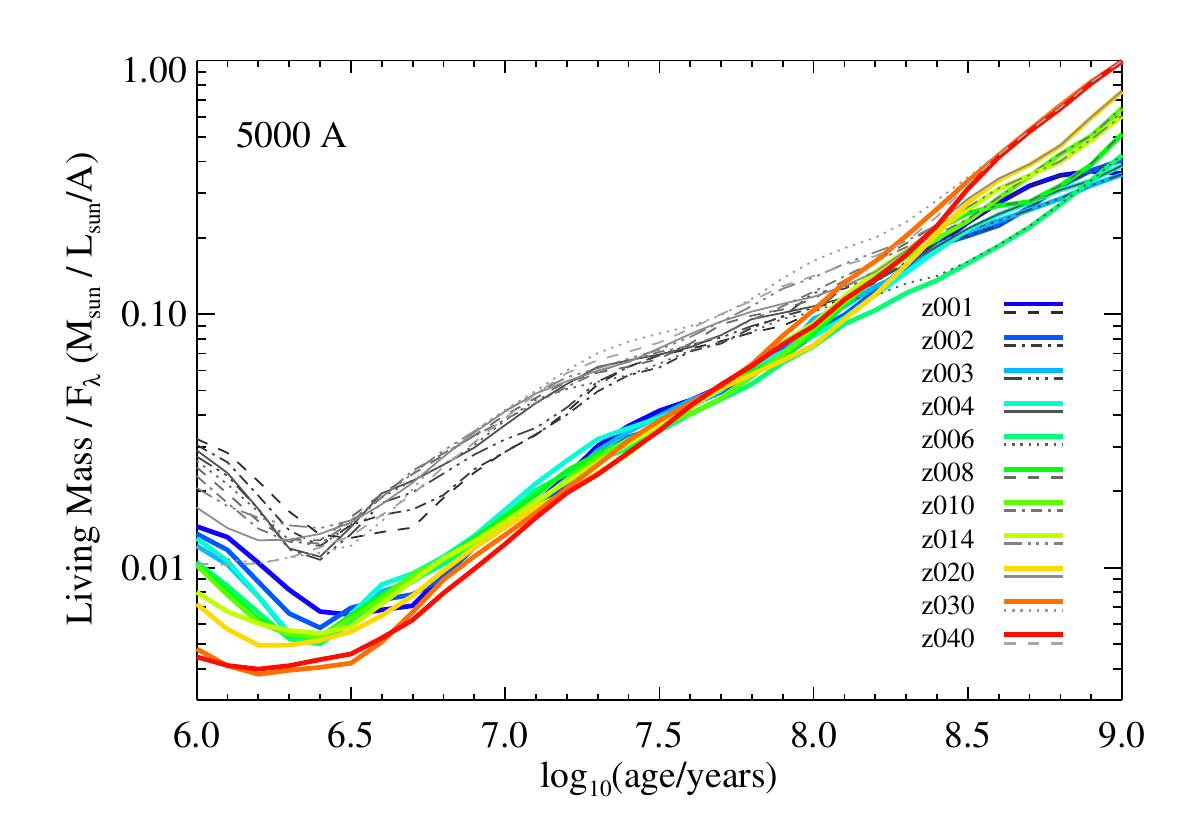}
    \includegraphics[width=0.98\columnwidth]{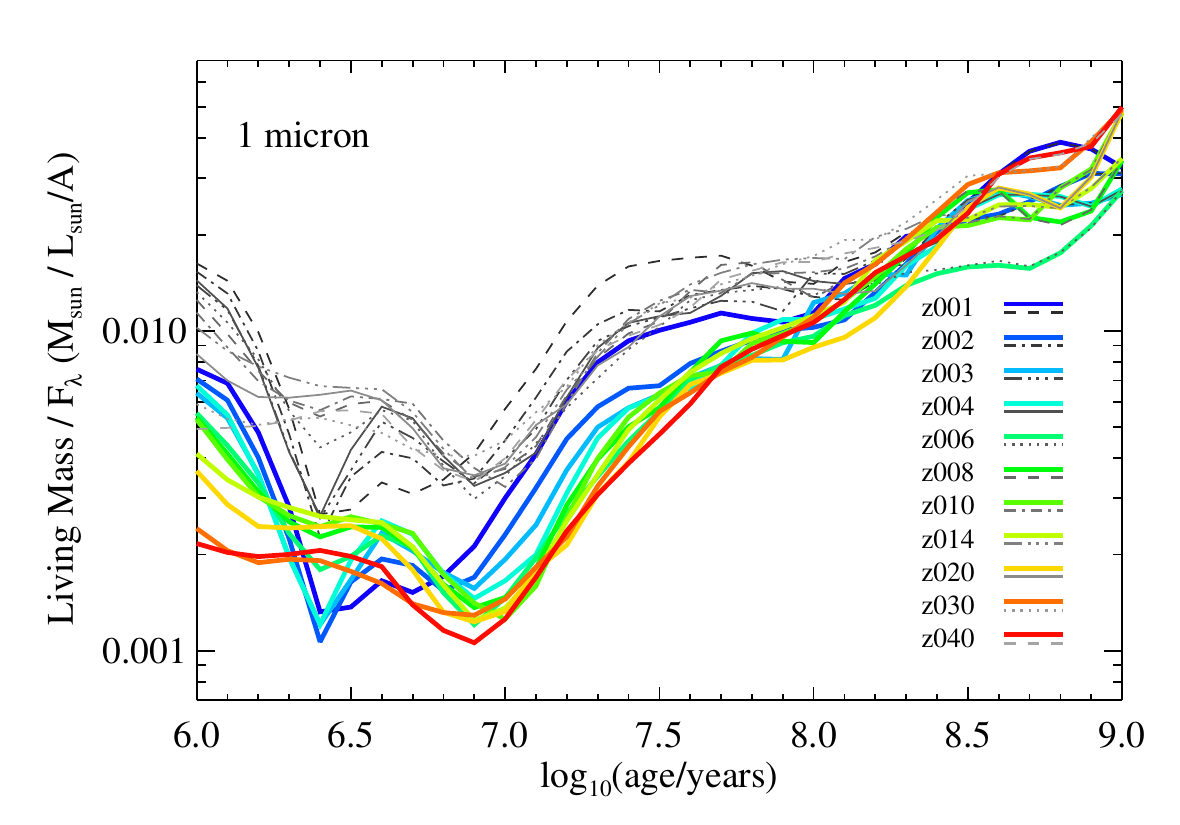}
    \caption{The impact of metallicity dependence of delay times and stellar lifetimes on the mass-to-light ratio. Values derived from a coeval ZAMS scenario are shown in greyscale, while those for a pre-MS delay scenario vary from red (Z=0.040) to blue (Z=0.001) as a function of metallicity.}
    \label{fig:mlratiowithZ}
\end{figure*}

In Fig.~\ref{fig:mlratiowithZ} we illustrate the impact of metallicity on the mass to light ratio in the optical and near infrared, in both the coeval ZAMS (greyscale) and pre-MS (colour) scenarios. High metallicity stars have longer pre-MS delay times in the \citet{2011A&A...533A.109T} prescription, but the same stars also exhibit stronger stellar winds in the BPASS evolution models. This results in a mass to light ratio that stays low for longer, but then evolves more rapidly than the equivalent at low metallicities. The prominent double-dip feature in the 1 micron mass to light ratio results from evolution of massive stars onto the giant branch at around log(age/years)=6.5 (which dominates at low metallicities), and then a further rise in luminosity at around log(age/years)=7.0 resulting from stellar atmosphere stripping as a combination of winds and binary interactions (which is strong at higher metallicities).

\subsection{Impact on Stellar Population Spectra}

\subsubsection{Ultraviolet spectra}\label{sec:res-uv}

\begin{figure}
    \centering
    \includegraphics[width=0.98\columnwidth]{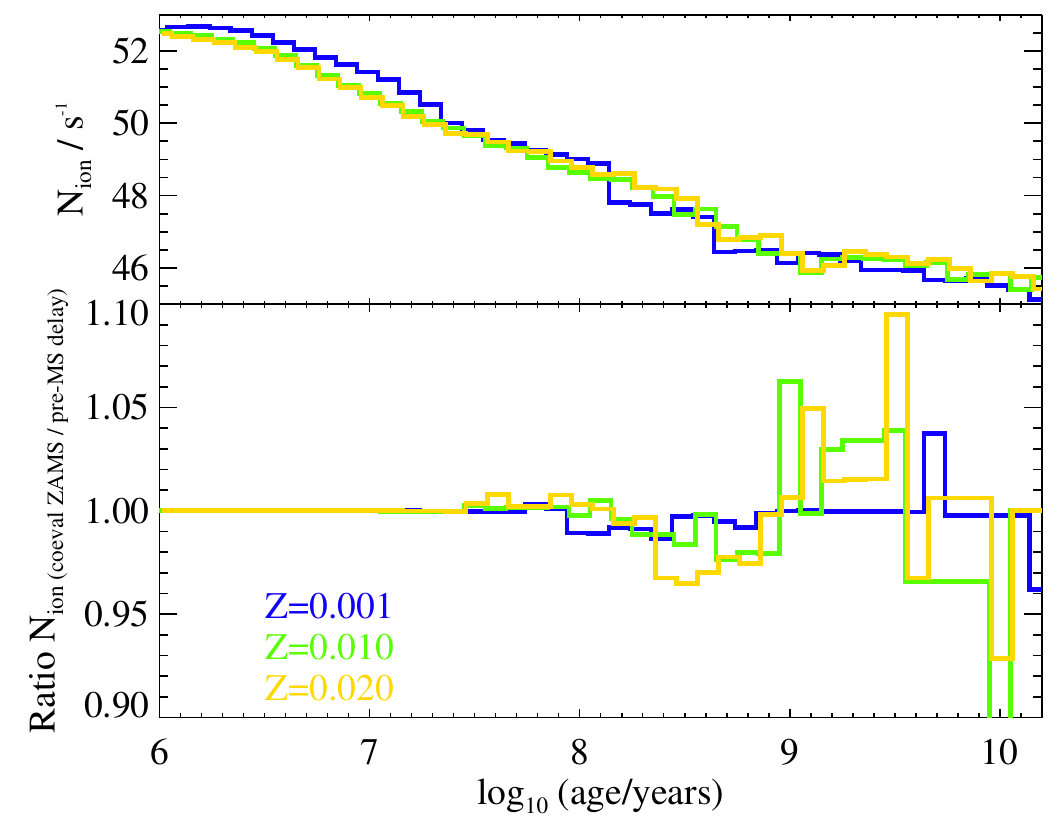}
    \caption{Metallicity dependence of the variation in ionizing photon production rate between the coeval ZAMS and pre-MS delay scenarios. We show lines at Z=0.001 (blue), Z=0.010 (green) and Z=0.020 (orange).}
    \label{fig:nion}
\end{figure}

\begin{figure}
    \centering
    \includegraphics[width=0.98\columnwidth]{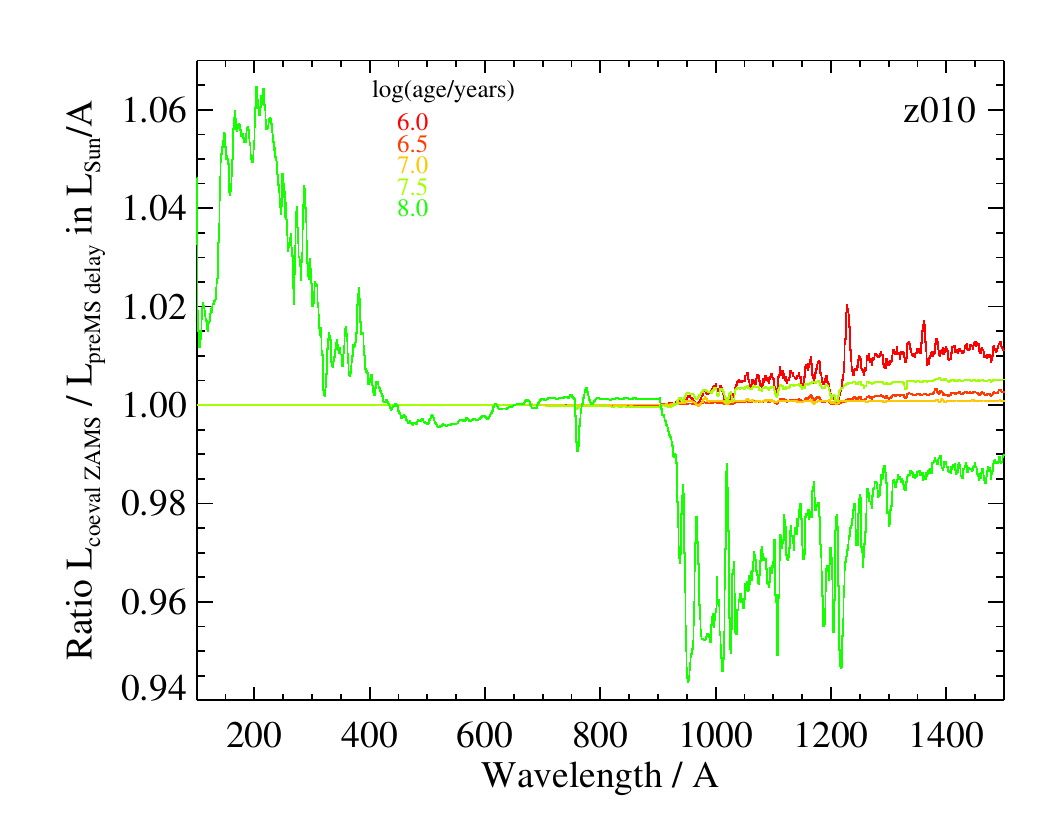}
    \caption{The variation in ultraviolet continuum spectra between the coeval ZAMS and pre-MS delay scenarios.}
    \label{fig:uvrat}
\end{figure}

The ionizing photon production rate (E$_\mathrm{phot}>$13.6\,eV) is entirely dominated by massive stars (M>10\,M$_\odot$) for which the pre-MS delay timescales are $<$1\,Myr and so have no significant effect, as demonstrated in Fig.~\ref{fig:nion}, until ages of around 25\,Myr, by which point most massive stars have died.  
 The details of the far ultraviolet spectrum are shown in Fig.~\ref{fig:uvrat}. As was the case for the ultraviolet continuum in Fig.~\ref{fig:mlratio}, there is a very small contribution to the ionizing photon production from stars in the initial mass range 1-5\,\msun\ as they are absent at zero age and mature towards stripped or partially stripped post-main sequence phases at slightly different epochs. As a result, the photon production rate shows variations of $\sim$2\,per cent between the coeval ZAMS and pre-MS delay scenarios at log(age/years)=8.5-9.5. However, by this population age the overall rate has declined by more than four orders of magnitude from its peak, and hence the impact is negligible. The same delay in the shift from main sequence to stripped stars manifests in a slightly larger difference in the ionizing spectrum at this epoch, but again the overall flux levels are extremely low.

The ultraviolet continuum flux density of a stellar population is often parameterised by a simple power law of the form F$_\lambda\propto\lambda^{-\alpha}$. Where spectroscopy of sufficient signal-to-noise is available, this can be measured directly through model fitting. However, given the relative expense of deep spectroscopy and the complex absorption and emission line blending in this part of the spectrum, the ultraviolet spectral slope, $\alpha$, is often approximated using a simple flux ratio derived from the photometric flux in available wavebands or from relatively line-free regions of the spectrum. 

\begin{figure}
    \centering
    \includegraphics[width=0.98\columnwidth]{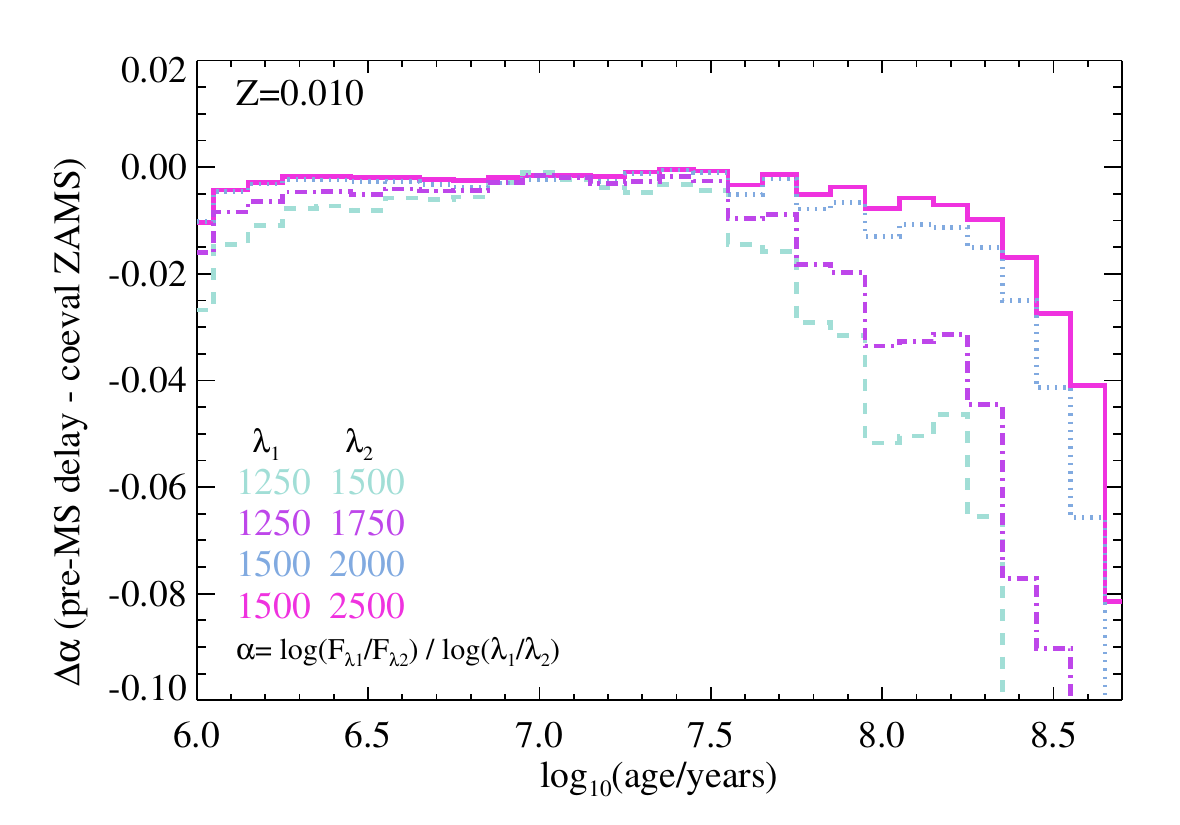}
    \includegraphics[width=0.98\columnwidth]{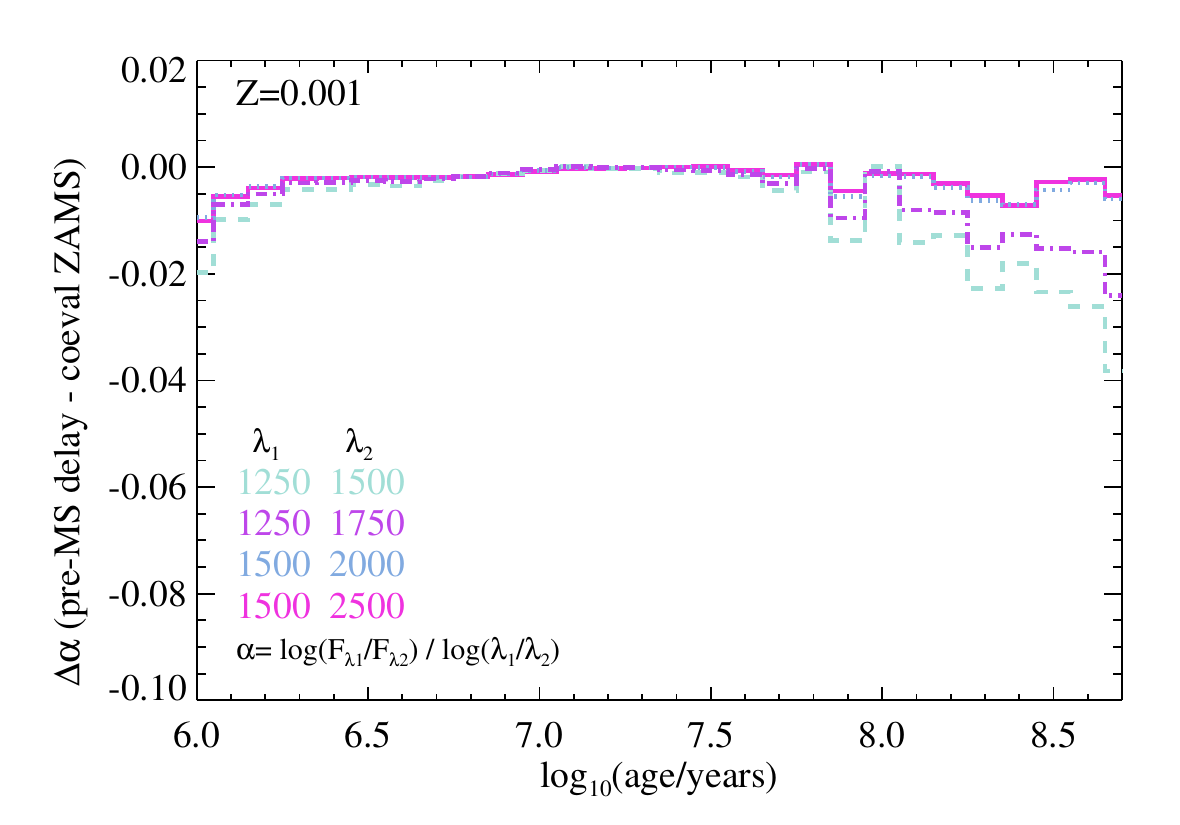}
    \caption{UV spectral slope, as derived from key continuum points}
    \label{fig:UVslope}
\end{figure}

In Fig.~\ref{fig:UVslope}, we demonstrate how the results derived using this approach are modified by the pre-MS delay scenario. We show results at two metallicities and with four different wavelength pairs in the far-ultraviolet. The change in power-law slope between the pre-MS delay and coeval ZAMS scenarios is typically less than $\Delta\alpha=0.02$, but increases rapidly with age beyond log(age/years)=7.5, particularly at higher metallicities. 

This difference is substantially less than the change in $\alpha$, often $\Delta\alpha>0.5$, that arises due to continuum emission from nebular gas irradiated by the stellar population, or the comparable changes that can arise from minor variations in star formation history. 

\subsubsection{Optical and Infrared spectra}\label{sec:res-opt}

\begin{figure}
    \centering
    \includegraphics[width=0.98\columnwidth]{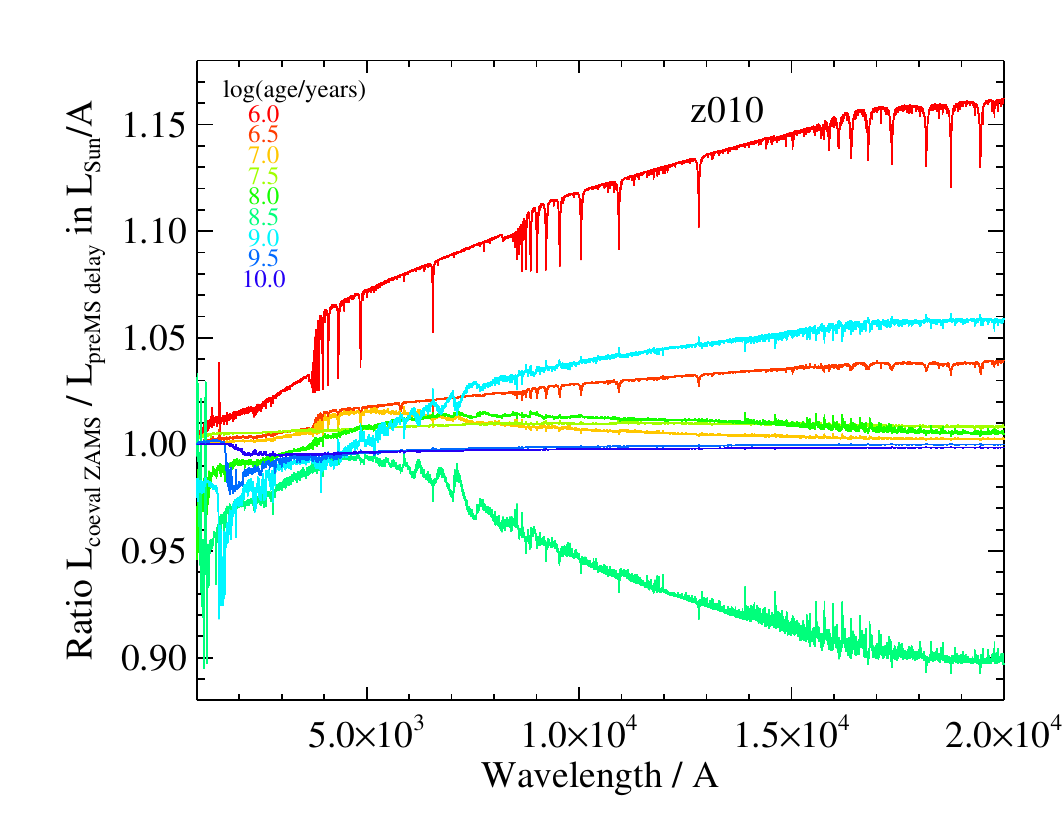}
    \caption{The variation in optical continuum spectra between the coeval ZAMS and pre-MS delay scenarios.}
    \label{fig:optrat}
\end{figure}

The impacts of pre-Main Sequence delays on the optical and into the near-infrared are much stronger than those in the ultraviolet. We demonstrate this in Fig.~\ref{fig:optrat}, which shows the ratio of simple stellar population spectra in the coeval ZAMS scenario to those when a pre-MS delay is included.

As the Figure illustrates, the stellar population at an age of 1\,Myr no longer includes emission from stars in the mass range 1-5\,\msun. While individually of moderate luminosity, these stars are plentiful compared to those at higher masses and so contribute 10-15\,per cent of the optical/Near-IR luminosity at this age in the co-eval ZAMS case.  The result is that the ratio between the spectra inferred in these two scenarios closely resembles the spectrum of a class A star, in particular showing an expected modification to the strong Balmer, Paschen and Brackett absorption lines. As would be expected, the luminosity ratio at intermediate ages (10-100\,Myr) is within 2 per cent of unity. However at ages of 300\,Myr to 1\,Gyr the different scenarios produce luminosities that fluctuate as stars in different mass ranges evolve with different timescales. At these ages, the pre-MS delay scenario leads to a spectrum in which late type K and M stars are younger and hotter, and hence produce more optical flux than in the coeval ZAMS case, and so the ratio shows an imprint of a source with the broad molecular absorption lines expected for such sources in the optical and a cooler spectrum through the near-infrared. Beyond ages of approx 3\,Gyr, this erratic behaviour again converges towards a ratio of unity, as the surviving stars are relatively low in mass and evolving more slowly.

\begin{figure*}
    \centering
    \includegraphics[width=0.95\columnwidth]{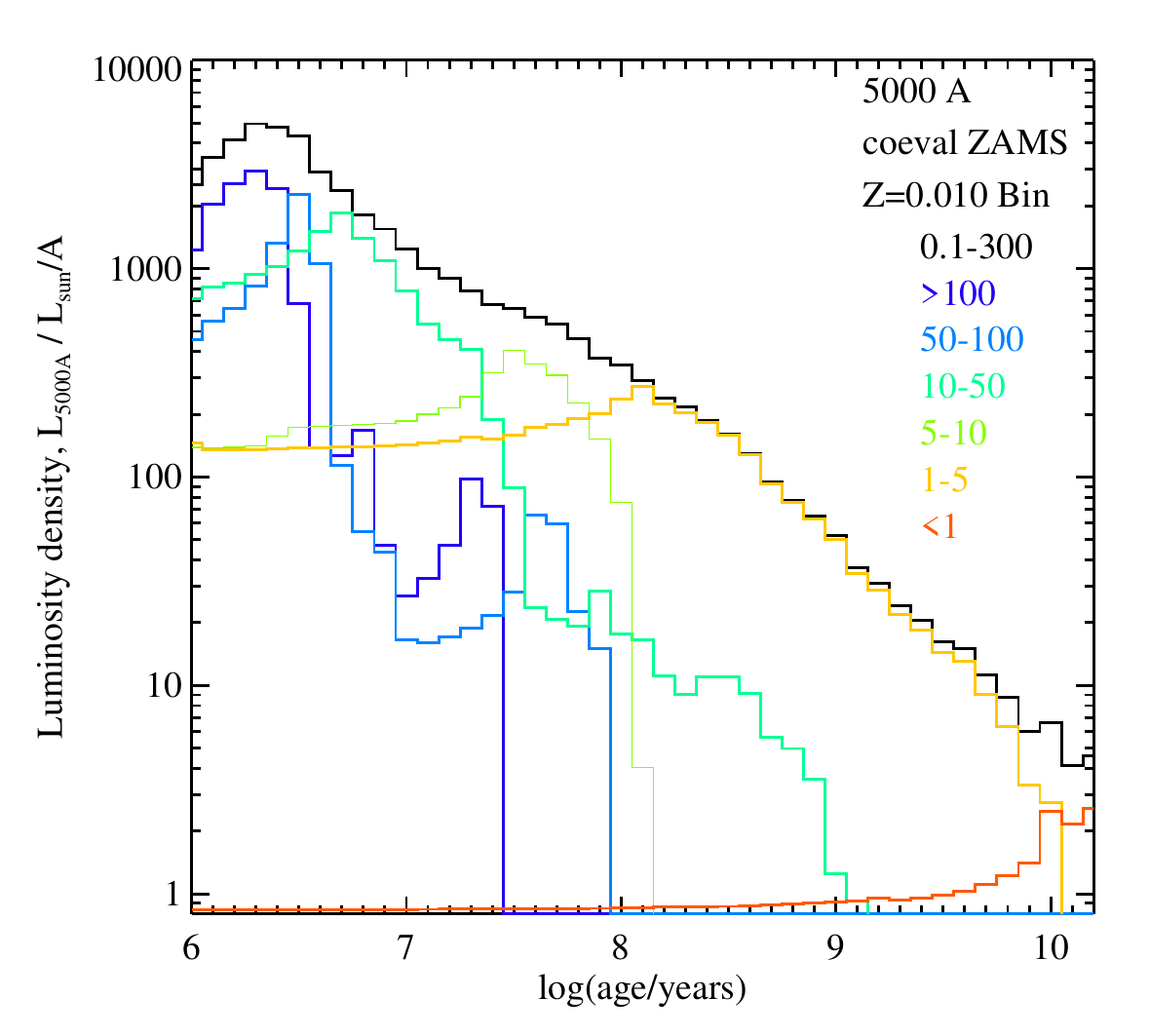}
    \includegraphics[width=0.95\columnwidth]{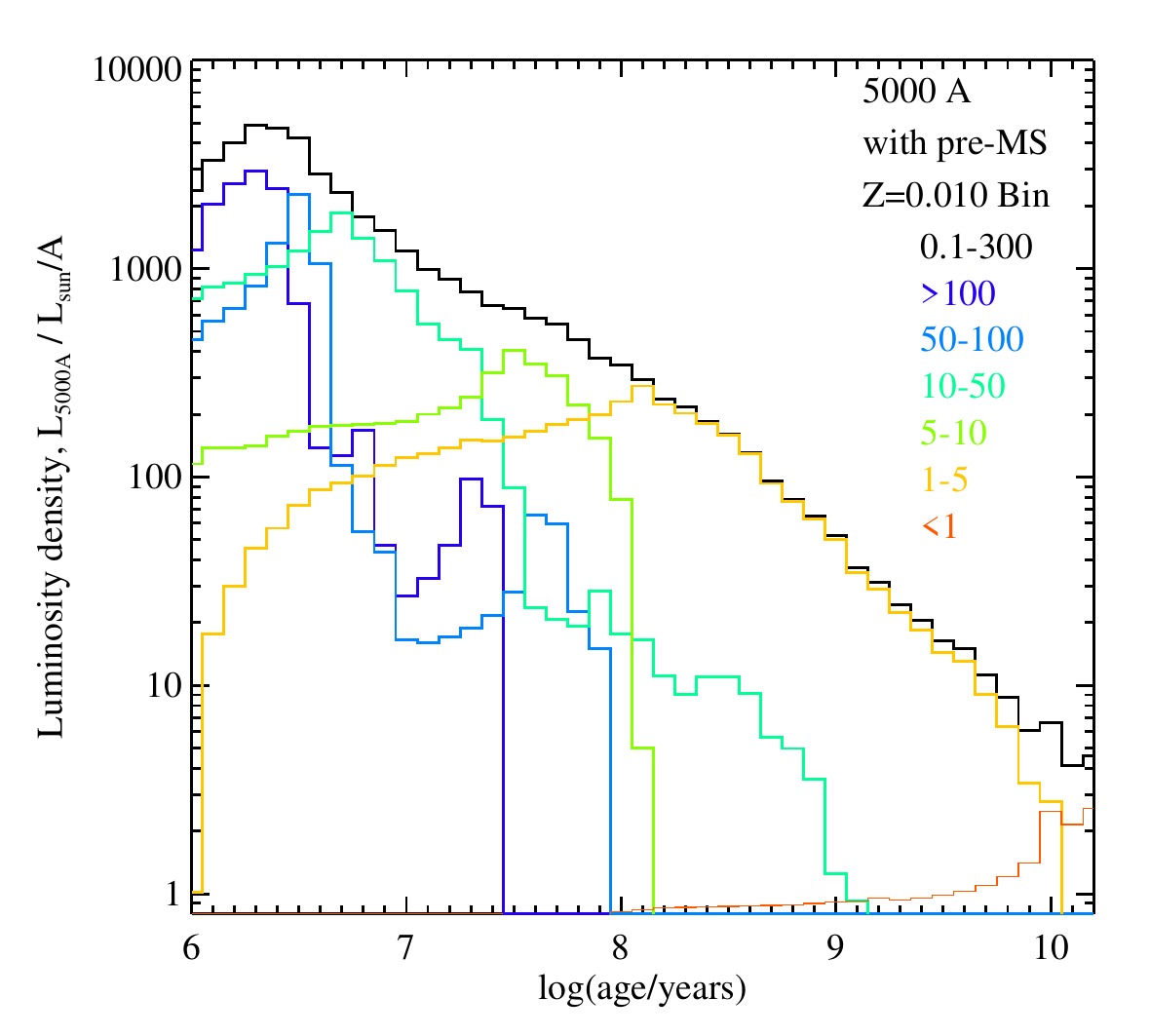}
    \includegraphics[width=0.95\columnwidth]{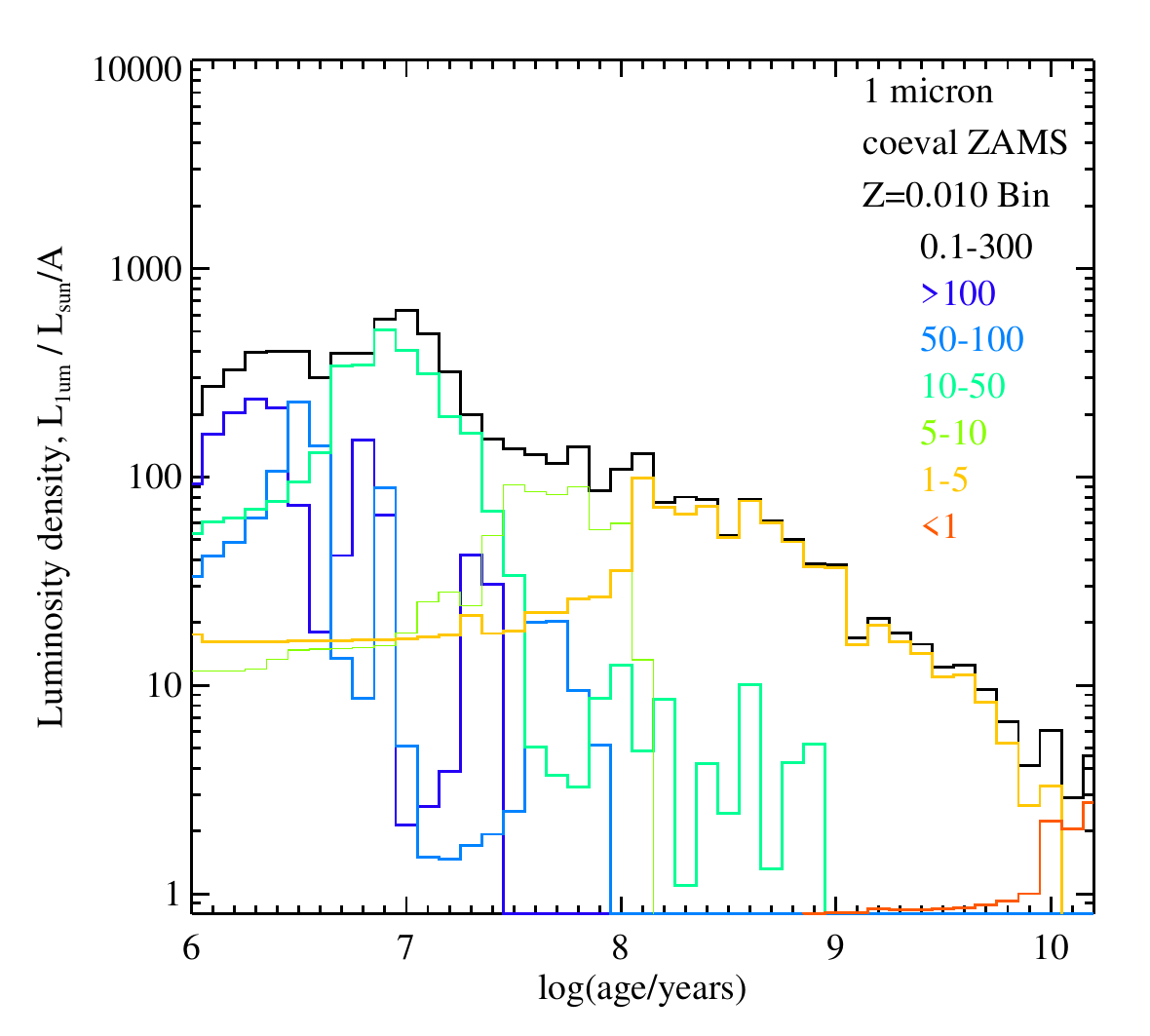}
    \includegraphics[width=0.95\columnwidth]{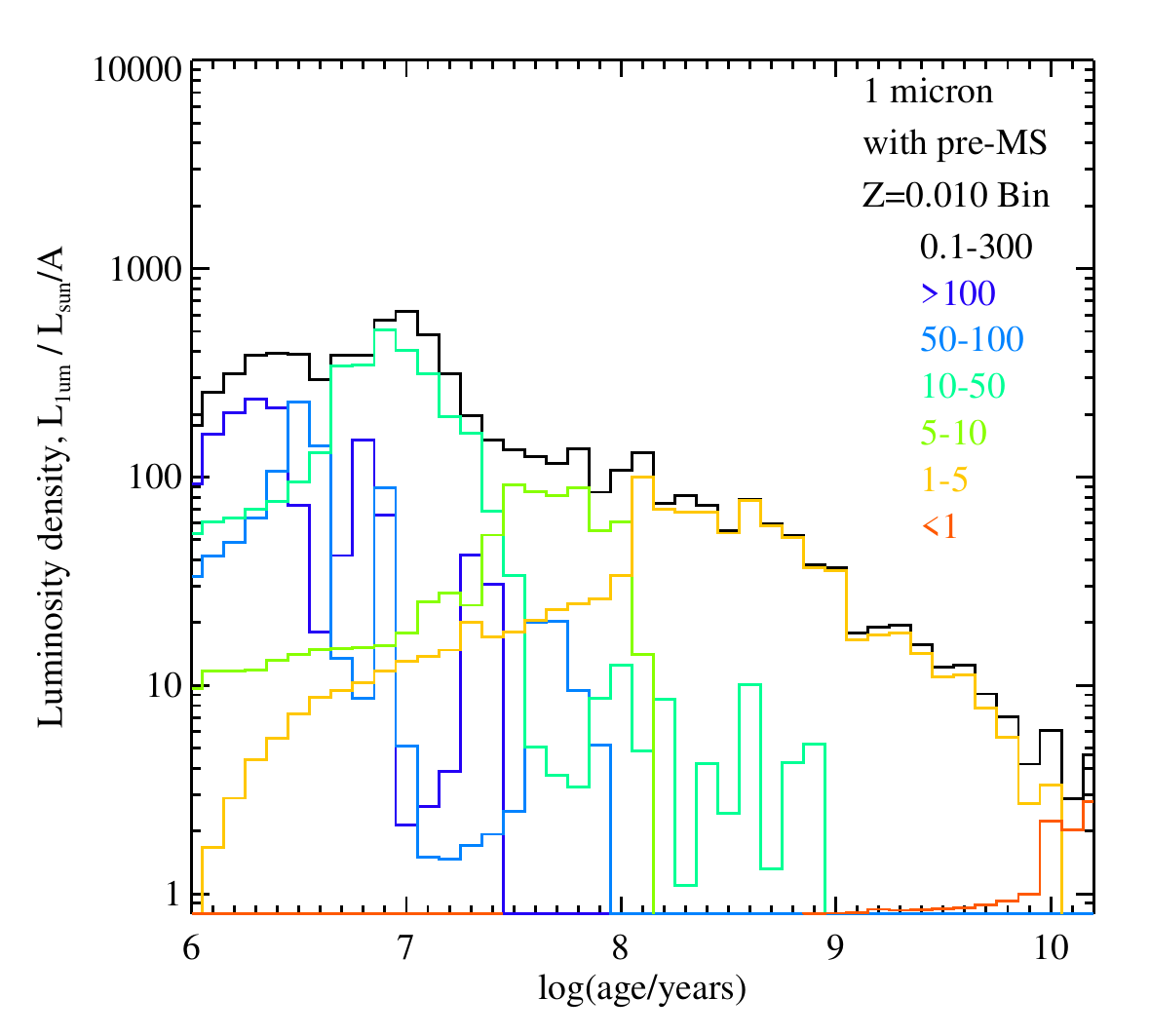}
    \caption{The time-dependent contribution to the total luminosity density (in L$_\odot$/\AA) of stellar populations with a final total stellar mass of $10^6$\,L$_\odot$ in different mass ranges at Z=0.010. The upper row gives contributions at a continuum wavelength of 5000\AA, while the lower row is the equivalent for 1$\mu$m. The left hand column is for the case of a coeval ZAMS epoch, while the right hand column incorporates pre-MS formation delays on low mass stars, with the total in the no-delay case shown as a dashed line. The population total (solid line) is divided to show contribution by primary mass into ranges \mzams$<$1 (orange), 1-10 (green), 10-100 (pale blue) and $>100$\,\msun (dark blue).}
    \label{fig:lumbymass}
\end{figure*}

As this figure makes clear, in the context of composite stellar populations and SED fitting, it is informative to consider the relative contributions of different stellar mass ranges to the population luminosity density, and how this varies as a function of time and wavelength.  In Figure \ref{fig:lumbymass} we calculate the contribution to the spectral energy density from stars in different ranges in primary (or single) star \mzams. In the case of binary systems, the secondary star flux is considered together with its primary rather than being assigned to a different mass bin.   The flux contribution assigned to massive and very massive stars continues to relatively late ages ($10-30$\,Myr) in these populations due to their binary construction (as they will have longer lived, less massive companions). 
This results from a combination of slightly lower mass companions being included with their primaries, and  primaries stripped by binary interaction which evolve on a longer timescale. 

As Figure \ref{fig:lumbymass} demonstrates, the optical continuum remains dominated by the most massive stars until population ages of 30-100\,Myr and is unlikely to be a robust indicator of stellar mass (rather than star formation rate) until that time. By contrast the 1\,$\mu$m continuum is relatively constant, with stars with \mzams$<10$\,\msun\ dominating at all ages and massive stars contributing less than ten per cent of the integrated light. This is true in both population age scenarios, although the impact of moderately luminous B and A stars is further delayed in the pre-MS formation case.

\section{Discussion}\label{sec:discussion}

\subsection{Interpreting Galaxy Candidates}

Recent years have seen a proliferation of candidate high redshift galaxies identified in JWST/NIRCAM data, as they were previously identified from HST/WFC3 and HST/ACS data at slightly lower redshifts. While an increasing number of such candidates now have JWST/NIRSPEC prism spectroscopy available to confirm their redshift, many do not, and full spectroscopic coverage is unlikely ever to be the norm for high redshift (in this context, $z>8$) galaxy samples. Thus we consider two possibilities: first that data on a target galaxy is only available from the NIRCAM photometric imaging, and secondly that this is complemented by NIRSPEC prism spectroscopy.

When a candidate is identified from photometry, several key items of information are typically measured. Amongst these, the most basic are the estimated location of spectral breaks (and hence  redshift), the UV photometric flux ratios and the UV-optical flux density. The photometric colours give rise to a joint estimation of the intrinsic UV spectral slope and dust extinction. The break location and flux density are used to estimate luminosity in typically between 3 and 5 photometric bands, and from these a joint estimate of star formation history, ongoing star formation rate and in-situ stellar mass can be made. To extract these quantities, the usual approach involves fitting the spectral energy distribution to models.  

When spectroscopy is also available, spectral line identification can lead to a more precise estimate of the systemic redshift. The luminosity of hydrogen recombination lines can be translated into an ionizing photon production rate, and thence into a more precise estimate of the star formation rate within the last 10\,Myr. The ratio of recombination lines is used to improve estimates of dust extinction, while other line ratios are further used to estimate nebular electron density and electron temperature, and thus metallicity and composition. The presence of high ionization emission and wind-broadened stellar absorption lines can also indicate the presence of unusual or extreme stellar populations. 

The majority of spectroscopically-derived measurements are likely unaffected by the inclusion or exclusion of the pre-main sequence evolutionary phase. As demonstrated in Figs.\,\ref{fig:nion} and \ref{fig:uvrat}, the ionizing photon production rate, ionizing spectrum and far-ultraviolet luminosity from a population differ in the two scenarios by no more than around one part in a hundred (although estimates of precision better than this might be called into question). The strongest deviation is in the Balmer lines at very early times, and these differ between the pre-MS delay and coeval ZAMS scenarios by no more than a few per cent, and over most of the evolutionary timescale by $<<$1 per cent. As a result, any emission that arises primarily from nebular gas shows negligible changes.

\begin{figure}
    \centering
    \includegraphics[width=0.98\columnwidth]{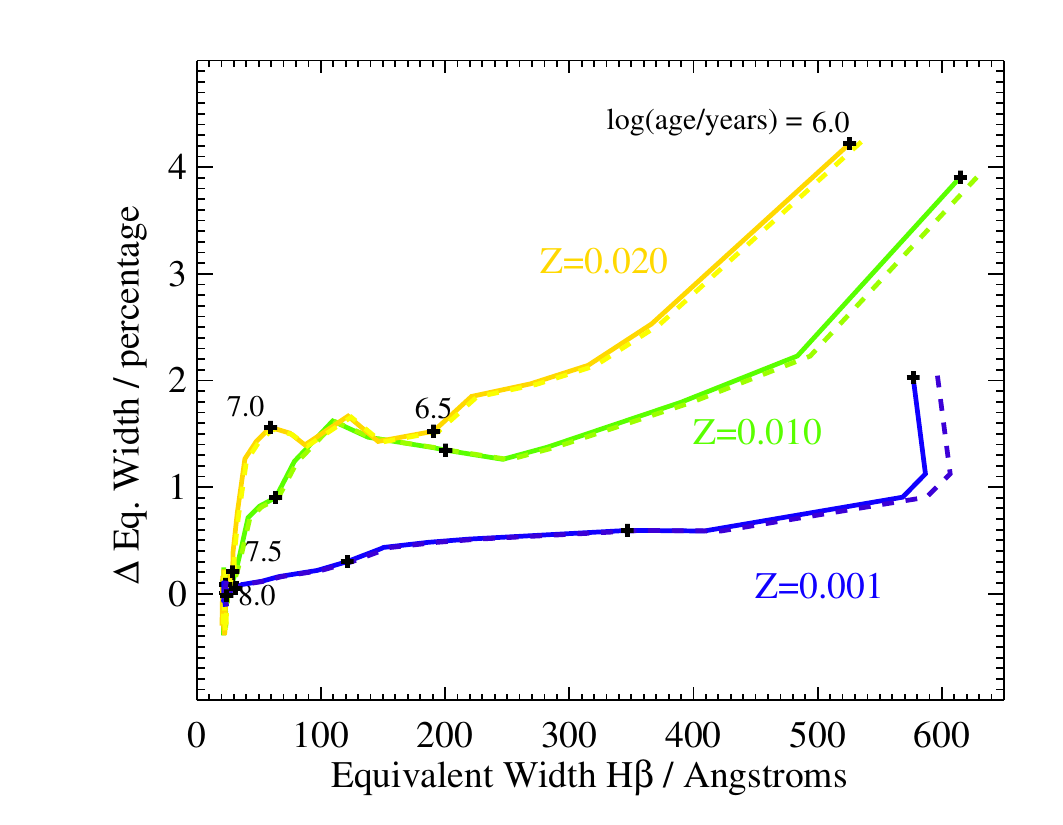}
    \caption{Metallicity dependence of the H$\beta$ line equivalent width ratio between the coeval ZAMS and pre-MS delay scenarios. We show lines at Z=0.001 (blue), Z=0.010 (green) and Z=0.020 (orange). The dashed line indicates the line EW derived when the continuum is accurately subtracted, while the solid line represents a simple summation of the flux across the line width.}
    \label{fig:HbetaEWwithZ}
\end{figure}

However any interpretation which depends on the behaviour of the stellar continuum, or its ratio to nebular emission components, presents a different challenge as demonstrated in Fig.~\ref{fig:HbetaEWwithZ}. 

Since line emission in young stellar populations is overwhelming produced by nebular gas, for the purposes of comparison, we process the simple stellar population models produced by BPASS with the radiative transfer code Cloudy \citep{2017RMxAA..53..385F}, assuming a fixed nebular gas electron density of $10^3$\,cm$^{-3}$ and gas cloud inner radius of 10\,pc (fairly typical for distant galaxies). Line emission is combined with the emergent stellar and nebular emission components at a resolving power of about 4500, with wavelengths evaluated in vacuum. Hydrogen recombination line features were visible in the spectral differences shown in Fig.~\ref{fig:optrat}, hence we expect these to be most sensitive to the pre-MS scenario. We focus on the H$\beta$ line at 4861.35\,\AA, rather than H$\alpha$ since the latter is often blended with the [N{\sc II}] doublet that straddles it.

We estimate line equivalent width in two ways. In each case the continuum flux density is inferred as the median flux density in a window defined by $20 < | \lambda - \lambda_\mathrm{line} | < 50$\,\AA. However solid lines indicate the equivalent width when line flux is estimated as $\Sigma\,F_\lambda ( |\lambda - \lambda_\mathrm{line} | < 15$\,\AA) in the output spectrum. In this case, no adjustment is made for any absorption feature underlying the emission line. Dashed lines indicate the equivalent when the stellar and nebular continuum (including any absorption features) are precisely subtracted and the emission line flux predicted by Cloudy alone is considered to contribute to the equivalent width. The latter case produces equivalent widths at early times that are slightly higher.

Despite the insensitivity of the ionizing flux to coeval ZAMS versus pre-MS delay scenario, the sensitivity of the optical continuum luminosity density results in a $\sim$4\,per cent offset in the H$\beta$ line equivalent width for stellar populations at 1 Myr, decreasing to typical values below 2\,per cent by the age of 2\,Myr, and decreasing further as the population ages. The offsets are smallest at low metallicities. We note that the line measurement technique has negligible effect on these offsets. 

Realistically, the difference in line equivalent width is small compared to uncertainties in observations in the distant Universe, even in a line such as this which contains conflicting components from both stellar and nebular sources. However JWST spectroscopy at $z=5-8$ is now bringing this order of precision within reach and the results here caution against over-interpretation of line widths without a careful consideration of all possible contributions to both ionizing flux and the optical continuum.

\subsection{Case Study: a high redshift galaxy candidate}
\label{sec:example}

To provide a more quantitative example, we consider the case of a simulated $z=10$ galaxy. We extract particle data for a galaxy selected at random from the high redshift suite of hydrodynamic simulations performed  by \citet{2020MNRAS.493.4315M,2019MNRAS.487.1844M} as part of the Feedback In Realistic Environments (FIRE) project \citep{2023ApJS..265...44W}. Galaxy FIRE2m11i has formed $1.25\times10^7$\,\msun\ of stars by the simulation snapshot at $z=10$, of which $1.7\times10^6$\,\msun\ has formed in the last 10\,Myr. The mass-weighted average age of star particles in the galaxy is log(age/years)=7.8 (60\,Myr). It has a fairly uniform metallicity of about $0.25$\,Z$_\odot$. Using their individual formation time and formation mass, we associate a BPASS stellar population at $Z=0.004$ with each star particle in the simulation. Hence we create a composite stellar population and stellar spectral energy distribution as would be observed from the light emitted at $z=10$.

The composite SED is redshifted to $z=10$, and its predicted spectrum calculated, assuming the source is observed with JWST/NIRSPEC, by smoothing to the appropriate wavelength-dependent resolution and then resampling onto the appropriate pixel dispersion scale. For dispersing elements we consider the prism, G140M and G395M gratings\footnote{Dispersion element characteristics were downloaded from \url{https://jwst-docs.stsci.edu/jwst-near-infrared-spectrograph/nirspec-instrumentation/nirspec-dispersers-and-filters}.}.

\begin{figure*}
    \centering
    \includegraphics[width=0.995\columnwidth]{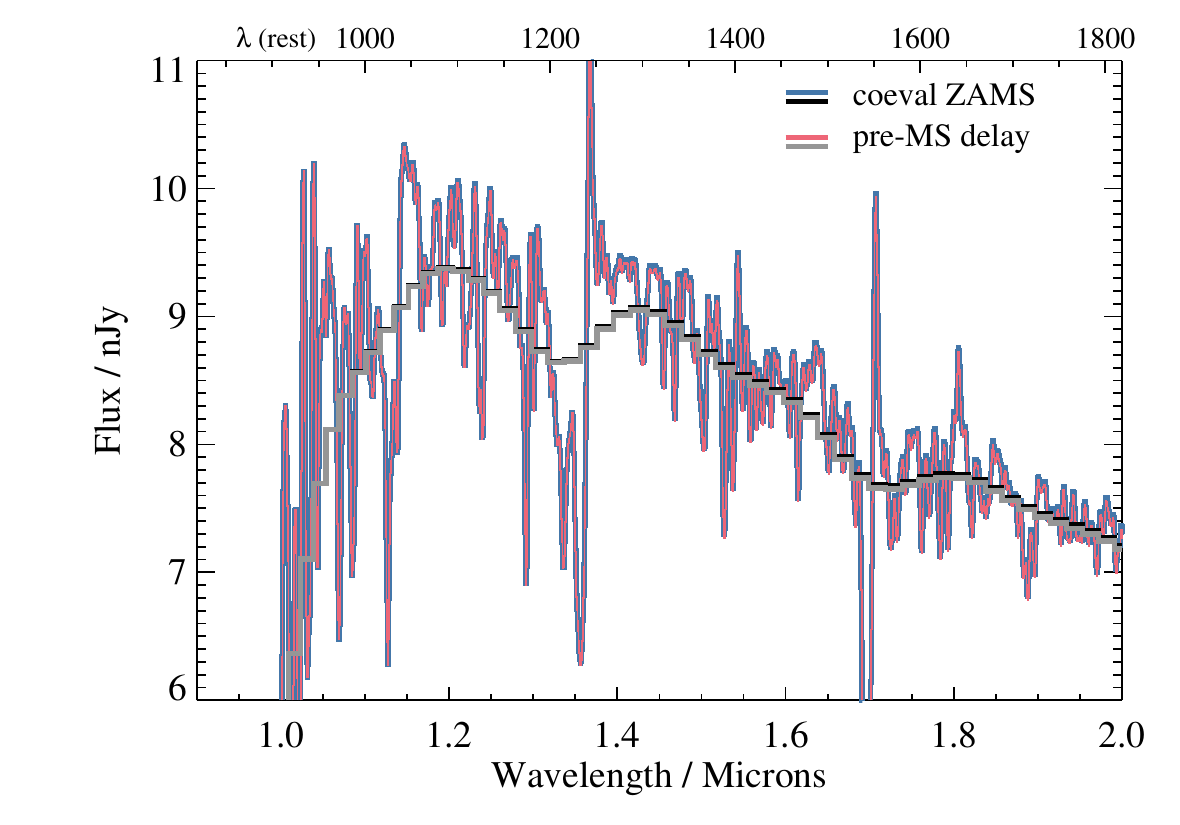}
    \includegraphics[width=0.995\columnwidth]{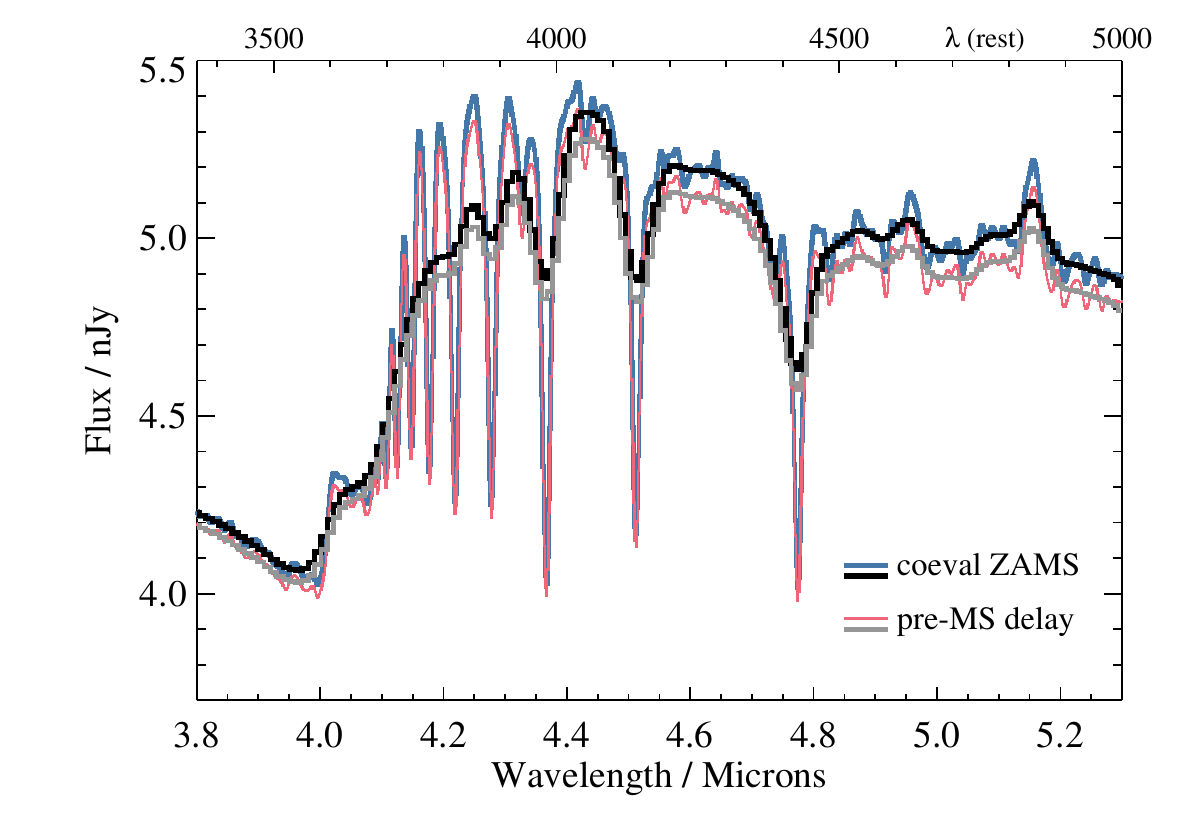}
    \caption{The contrast between simulated stellar emission spectra for $z=10$ simulated galaxy FIRE2m11i in the pre-MS delay and coeval ZAMS case. On the left, we show the rest-frame far-ultraviolet as would be observed using the JWST/NIRSPEC G140M grating and by the NIRSPEC prism. On the right, we show the rest-frame near-UV and optical as visible to the JWST/NIRSPEC G395M and prism dispersing elements. In each case, the prism is the lower resolution spectrum.}
    \label{fig:example_stellarspec}
\end{figure*}

The resultant model spectra are shown in Fig.~\ref{fig:example_stellarspec}. As expected, for a galaxy with a substantial young stellar population, the rest-frame far-ultraviolet shows no significant difference between the coeval ZAMS and pre-MS delay scenarios. The NIRSPEC instrument extends to 5.3\,$\mu$m, just past the near-UV/optical boundary. In this region of the spectrum, the stellar flux of FIRE2m11i at $z=10$ typically differs by about 0.1 nJy on the 5 nJy continuum, or two percent, consistent with expectations for a stellar population with a typical age of a few Myr (as shown in Fig.~\ref{fig:optrat}) (i.e. the light is dominated by the youngest stars). 

\begin{figure*} 
    \centering
    \includegraphics[width=0.995\columnwidth]{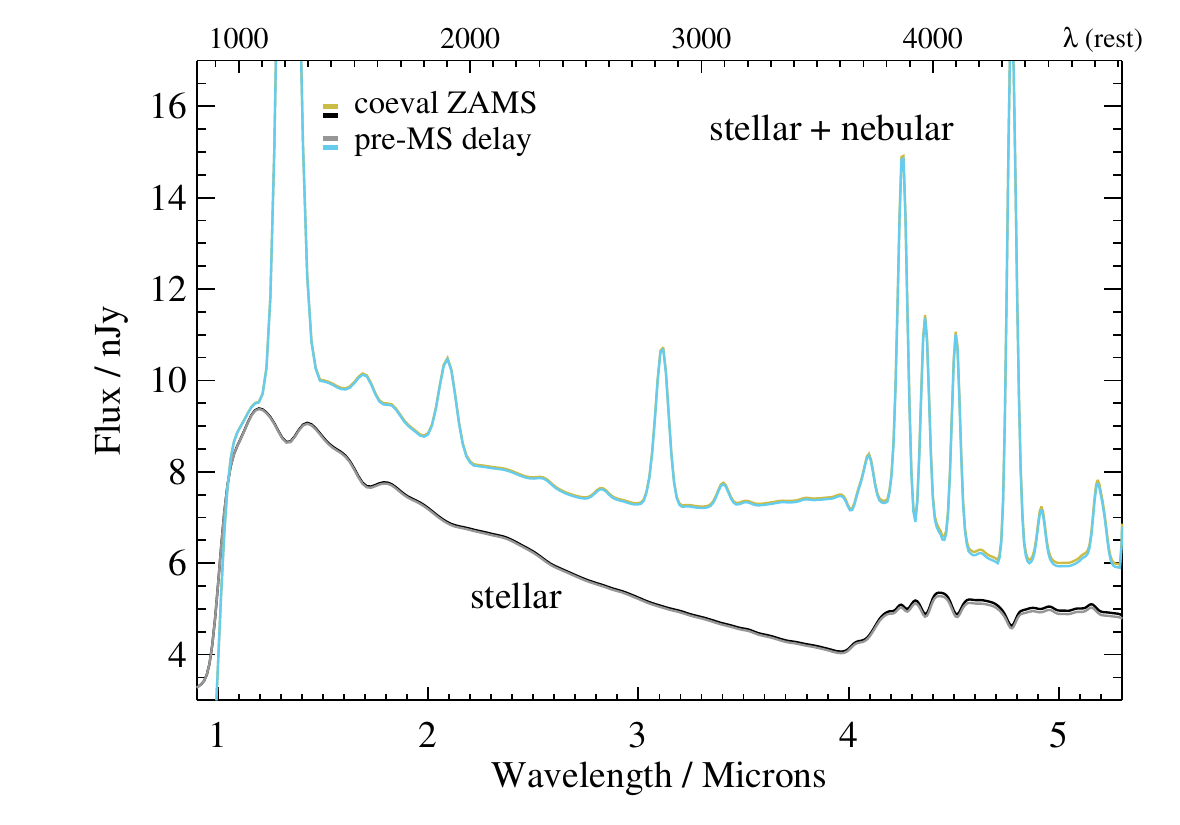}
    \includegraphics[width=0.995\columnwidth]{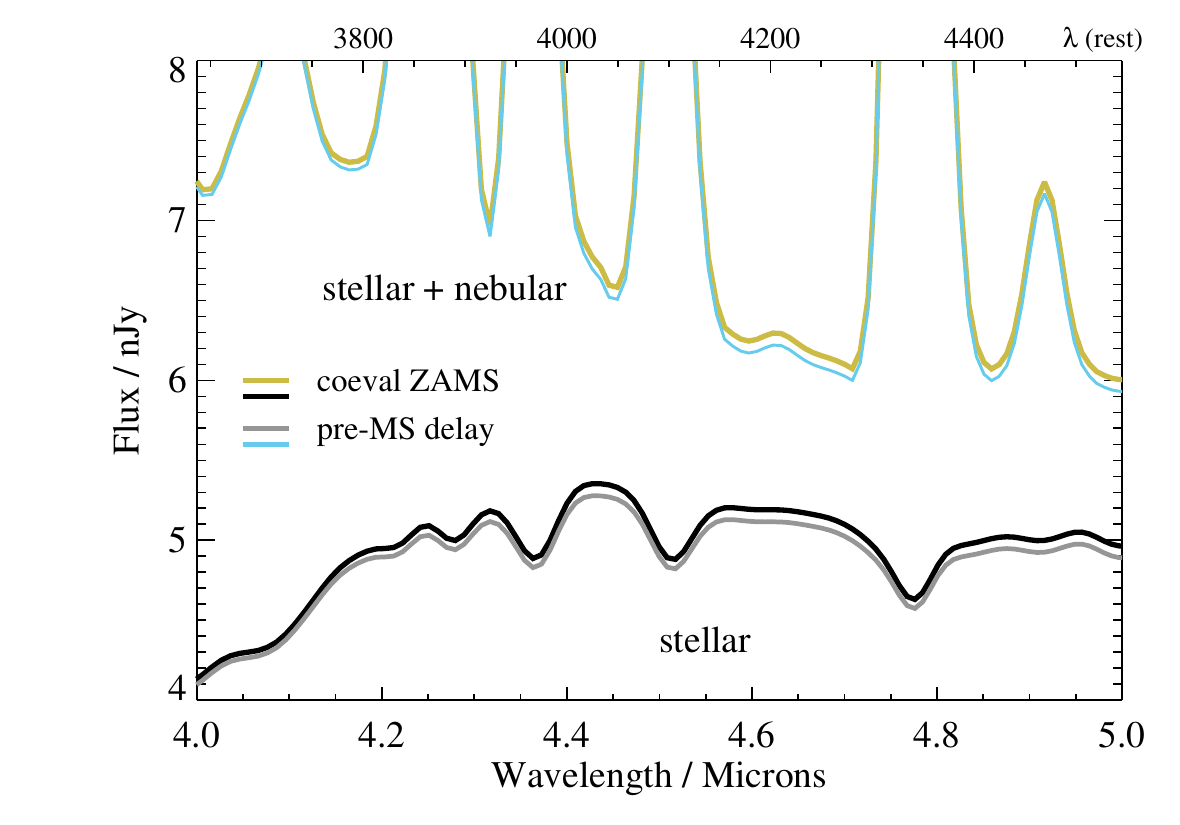}
    \caption{The contrast between simulated stellar emission and stellar plus nebular emission spectra for $z=10$ simulated galaxy FIRE2m11i in the pre-MS delay and coeval ZAMS case. On the left, we show the full wavelength range observed by the NIRSPEC prism. On the right, we focus on the rest-frame optical continuum in order to demonstrate the impact of the different scenarios.}
    \label{fig:example_stellarspec2}
\end{figure*}

Since this is a star forming galaxy, and hence gas-rich, we also calculate the equivalent composite populations including a nebular gas emission component. For this we again use the radiative transfer code Cloudy v17.03 \citep{2017RMxAA..53..385F}, matching the gas composition to that of the stellar models, and assuming spherical geometry with an inner gas cloud radius of 10\,pc and an electron density of 1000\,cm$^3$. Emission lines are assumed to have a velocity width of 100\,km\,s$^{-1}$ (although at this redshift such a line is typically unresolved). Again, we convolve the resulting spectra to the expected response of the JWST/NIRSPEC prism and show the output spectrum in Fig.~\ref{fig:example_stellarspec2}.

While the stellar continuum in the rest-frame near-UV and optical showed the expected impact of missing starlight in the pre-MS delay scenario, the apparent spectrum of this galaxy in NIRSPEC data remains dominated by nebular line emission (much of it highly blended). The strength of the nebular continuum is demonstrated by the clearly apparent Balmer jump, or increase in flux shortward (rather than longward) of the 4000\AA\ break. The nebular continuum continues to contribute twenty to forty per cent of the total continuum emission out to the 5\,$\mu$m limit of the spectrograph. As a result, the impact of the pre-MS delay scenario on the integrated light of the population remains small (of order 0.3 per cent or less across the observed spectral window), and is only plausibly detectable in the stellar continuum if the nebular conditions are extremely well constrained (beyond the scope of current facilities at this redshift). 

\begin{figure}
    \centering
    \includegraphics[width=0.98\columnwidth]{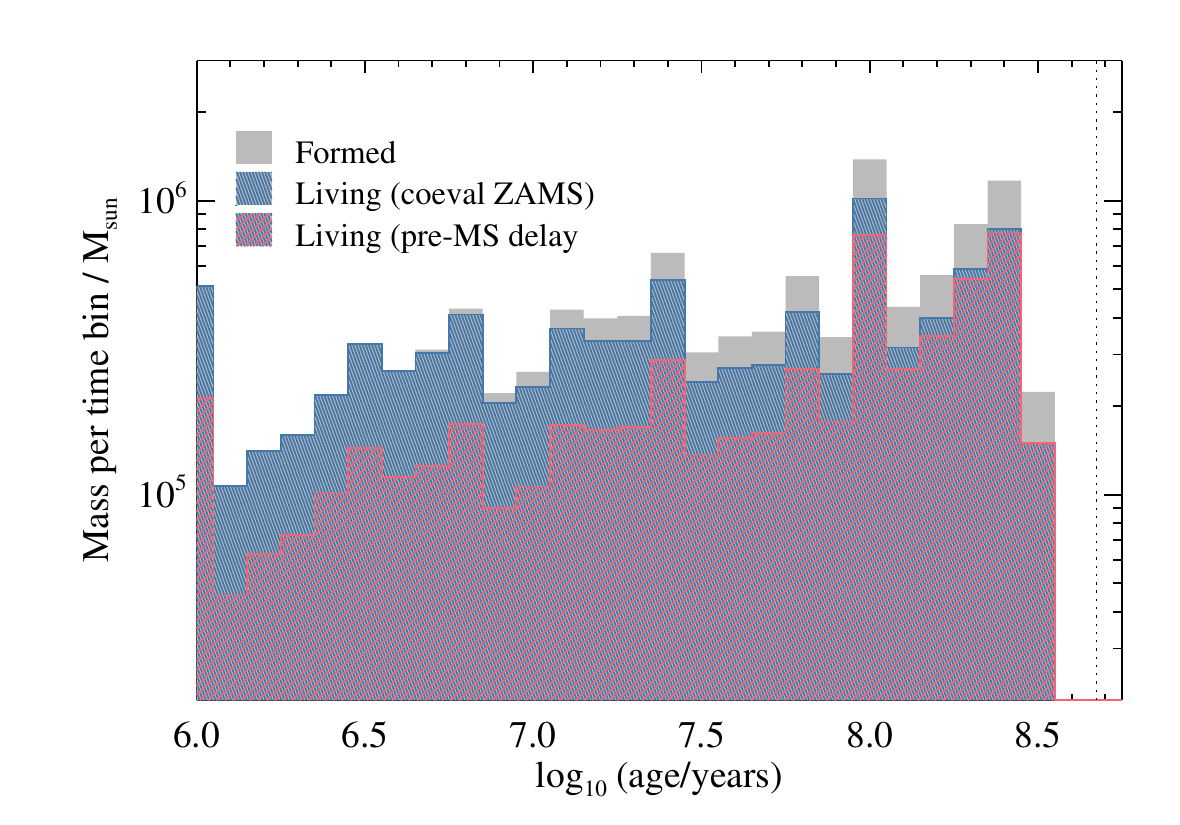}
    \caption{Mass formation history of a simulated $z=10$ galaxy FIRE2m11i, as described in section \ref{sec:example}. The grey bars indicate the formation history of stars in FIREm11i, while the blue and pink bars indicate the mass of living stars in the coeval ZAMS and pre-MS delay scenarios. Stars are sorted by time elapsed since the onset of star formation. The age of the Universe at $z=10$ is indicated by the vertical dotted line.}
    \label{fig:examplemass}
\end{figure}

By contrast, Fig.~\ref{fig:examplemass} shows that the interpretation of the detected light is different in the two scenarios. Of the $1.25\times10^7$\,\msun\ of stars formed by the simulation snapshot before $z=10$, $2.87\times10^6$\,\msun\ (22.9\,per cent) would have reached the end of its nuclear burning lifetime before the point of observation under both the coeval ZAMS and pre-MS delay scenarios. However under the pre-MS delay scenario, another $2.86\times10^6$\,\msun\ (22.8\,per cent)  would  exist in the form of proto-stars, yet to reach hydrogen ignition in their cores.

\begin{figure}
    \centering
    \includegraphics[width=0.98\columnwidth]{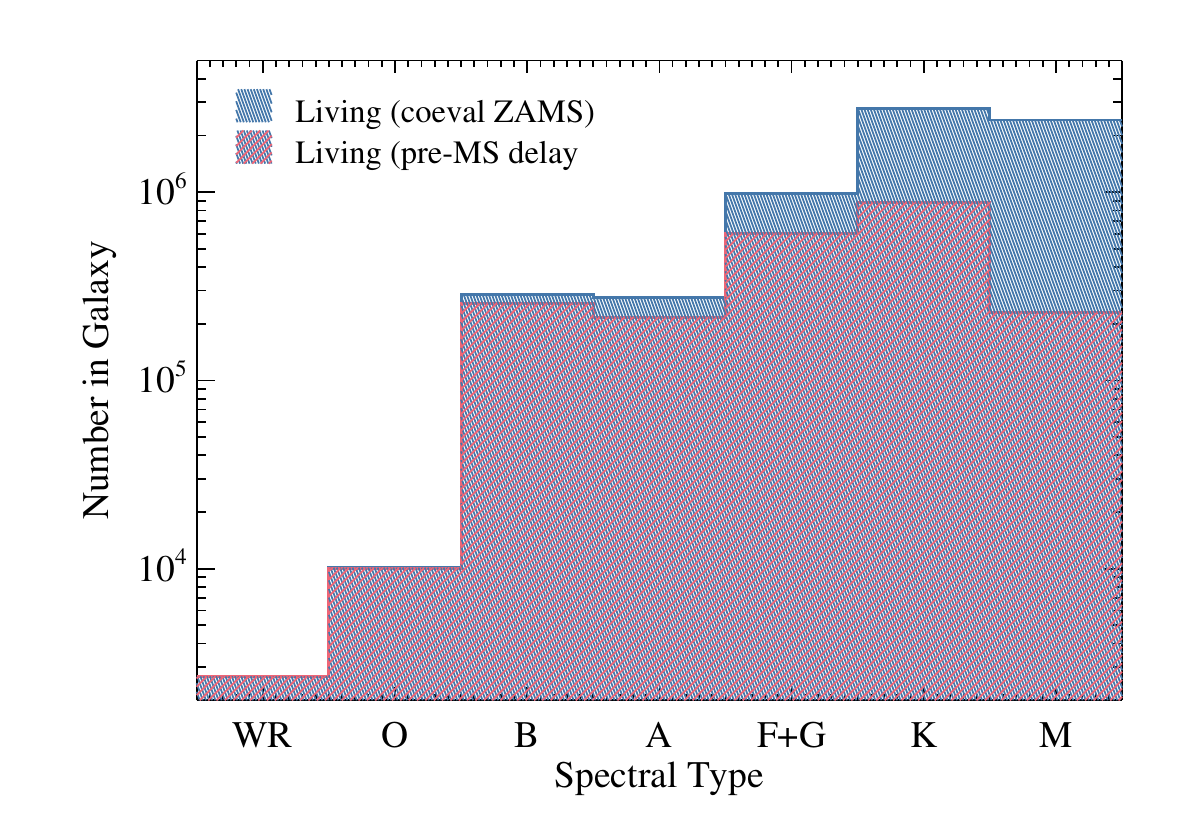}
    \caption{The number of nuclear burning stars as a function of spectral class in simulated $z=10$ galaxy FIRE2m11i, as described in section \ref{sec:example}. }
    \label{fig:exampletype}
\end{figure}

Crucially, both these fractions will also depend on the distribution of masses of stars, and hence the stellar initial mass function, in different ways. The protostar mass fraction will be higher for populations in which the IMF favours low mass stars, while the dead star fraction will be higher for IMFs rich in high mass stars. Since neither population is emitting detectable light (short of the far-infrared), their contribution to the mass budget can only be estimated. The distribution of stellar types amongst living stars in Galaxy FIRE2m11i is shown in Fig.~\ref{fig:exampletype} and confirms that the stars missing in the pre-MS delay scenario are, as expected, primarily of late spectral types, although a handful of early type stars that result from binary interactions are also absent from the observable population.

\subsection{Stellar Mass and Star Formation Rate}

So is the stellar mass of the simulated galaxy FIRE2m11i best described as M$_\ast=1.3\times10^7$\,\msun\ (since this is the total mass of the stellar populations represented by living members), M$_\ast=9.7\times10^6$\,\msun\ (the total mass of stellar populations whose formation has begun, not including stars that have already died) or M$_\ast=6.8\times10^6$\,\msun\ (the total mass of stars emitting light through nuclear burning)? While this might be regarded as a philosophical question, we note that the plausible range of values spans 0.27\,dex - significantly more than the uncertainties quoted on typical stellar masses.

\begin{figure}
    \centering
    \includegraphics[width=0.98\columnwidth]{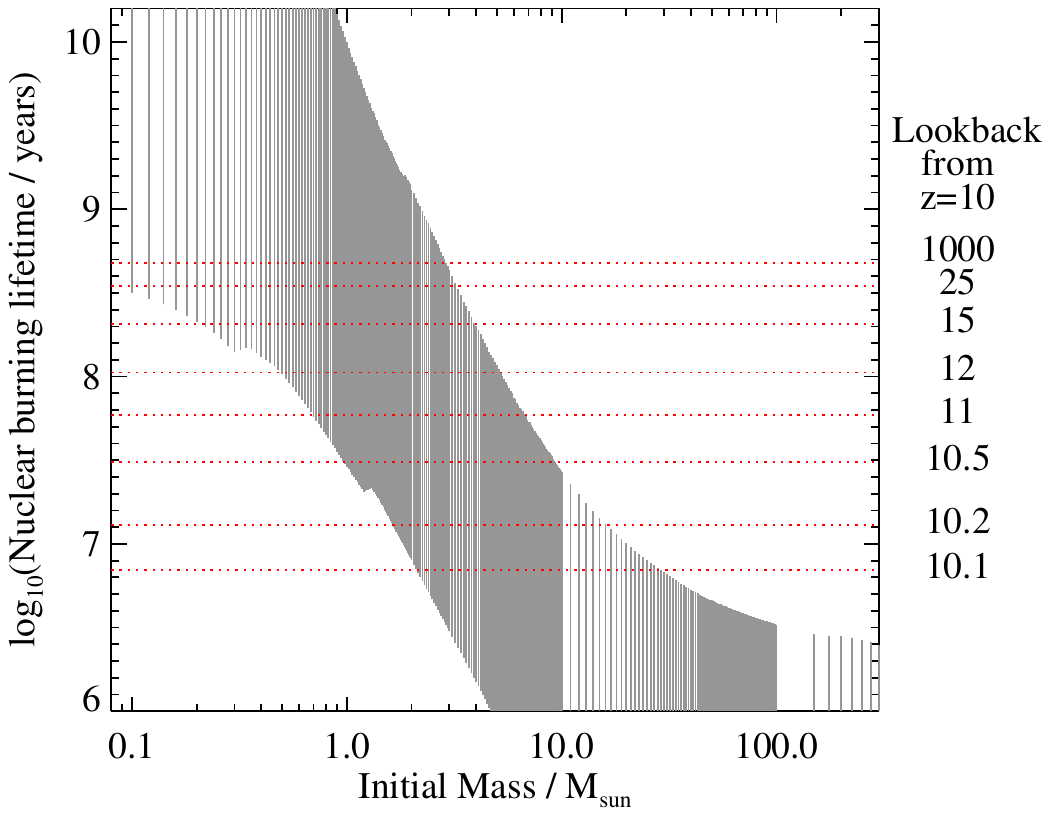}
    \caption{The nuclear burning timescales for stars as a function or mass in the pre-MS delay scenario, together with the equivalent formation redshifts for a galaxy observed at $z=10$. Each vertical line indicates the nuclear burning lifetime of an individual single stellar evolution model in the BPASS Z=0.010 grid.}
    \label{fig:examplelifetimes}
\end{figure}

Of course, the impact of pre-main sequence on stellar clusters in the local Universe is well known.  The nuclear burning lifetimes of massive stars and low mass stars in the same population are unlikely ever to overlap, as shown in  Fig.~\ref{fig:examplelifetimes}, and this leads to a fundamental inability to measure the full initial mass function of a stellar cluster in formation \citep[e.g.][]{2024ARA&A..62...63H}. A widely adopted convention for giving the stellar mass of a cluster is to quote the total initial mass of a population at ZAMS, and to include proto-stars and no-longer-extant starts in that mass budget \citep[e.g.][]{2019ARA&A..57..227K} or alternatively to quote the total of the stellar and gas mass of the cluster \citep[e.g.][]{2010ARA&A..48..431P}. However in most studies of regimes where this has a significant impact, the population is resolved and the mass of each star individually quantifiable.  

By contrast the stellar masses quoted for galaxies are usually given as the estimated contemporaneous stellar mass at time of observation, excluding stars which have already ended nuclear burning. Synthetic spectral energy distributions are corrected for these expired stars before the stellar mass is determined. In the relatively local Universe, where the mass (if not necessarily the light) is entirely dominated by populations exceeding a gigayear in age, this presents no real difficulty. 

As we have demonstrated here, in the distant Universe the situation is less clear cut.  The horizontal lines on Fig.~\ref{fig:examplelifetimes} indicate the maximum available time for stellar evolution since a range of plausible redshifts, for a galaxy such as our simulated example at $z=10$. Even stellar populations which began forming at the highest plausible redshifts, between $z=$15 and 25, would not have been in existence sufficiently long for the full stellar population to have reached the onset of nuclear burning.

We note in passing that a similar philosophical question could be applied to the concept of instantaneous star formation rate, although here the definition is somewhat clearer. A star formation rate of 1\,M$_\odot$\,yr$^{-1}$ is applied in all contexts to mean that a population of protostars of total mass $10^6$\,M$_\odot$  will have begun their journey towards the main sequence within the last 1\,Myr. Since the timescales for massive star formation are short, and the light in such young populations will always be dominated by the most massive stars (Fig.~\ref{fig:lumbymass}), the absence of nuclear burning in the low mass stellar population does not affect the interpretation of the ultraviolet luminosity, ionizing photon flux or its proxies, as a star formation rate. Where caution must be applied is in interpreting the integral of the star formation rate over time as the in-situ stellar mass.  For populations with typical ages of less than a gigayear, this will only be true if protostars and stellar remnants are included in the mass budget, and care must also be taken to account for the death of high mass stars and potential return of material to the interstellar medium by stellar winds and supernovae.

\subsection{Pre-MS interactions with Multiplicity and IMF}

In the preceding, we have assumed throughout that the secondary star in a binary has the same pre-main sequence lifetime as its primary, and that in both cases these match the pre-main sequence lifetime of a single star of equivalent mass to that primary. However, there are reasons to question that assumption. As we noted when describing our procedure in section \ref{sec:bpass_sps}, the role of feedback from stellar neighbours in shaping a star's pre-main sequence evolution is not well understood, particularly at the low metallicities prevalent in the distant Universe. 

The more massive stars, which are assumed to reach the main-sequence first, will provide feedback in terms of radiation and stellar winds. This feedback affects the newly forming and evolving stars both negatively and positively. Star formation may be suppressed, delayed or enhanced depending on the local gas density and heating, and resulting turbulence, since these affect the timescales for gas infall and accretion and may channel material onto protostars or result in fragmentation \citep[e.g.][]{2009Sci...323..754K,2018A&A...616A.101K,2024arXiv241212809N}. As a result, stellar feedback may help shape the low mass end of the stellar IMF, and also modify the pre-main sequence lifetime of low mass stars - either those forming as part of a dense cluster or in a gravitationally-bound multiple system. 

Given the relative scarcity of high mass star forming regions in the Local Group, together with the necessity of infrared data to explore dust-embedded proto-stellar populations, observational constraints on such processes are limited. The best local laboratory for studying these processes is arguably the 30 Doradus star-forming region in the Large Magellanic Cloud, also known as the Tarantula Nebula, which is forming stars intensively at a metallicity of around 0.5\,Z$_\odot$. This region has been intensively studied in recent years, at least in part in its role as a potential analogue to galaxy-wide starbursts in the more distant Universe \citep[see][]{2024MNRAS.527.9023C}. 

\citet{2016ApJS..222...11S} compiled a catalogue of more than 800,000 photometric sources reaching a mass limit of 0.5\,\msun\ in the Tarantula Nebula. By including near-infrared data, they were able to identify and catalog pre-main sequence candidates as well as individual massive stars. This showed clear evidence for low-mass star formation being suppressed in the hot ionized bubbles immediately adjacent to massive stars, but also of it being enhanced in the cooler shells of material swept up or disrupted by their winds. These effects may negate one another on global scales, and likely both contribute to the range of empirical initial mass functions observed in known stellar populations \citep[e.g.][]{2018PASA...35...39H}.

However such radiative feedback may have a larger impact by the ongoing formation and accretion of protostar \textit{companions} to massive stars. To some extent, this is accounted for in the empirical binary mass ratio and period distributions implemented for binary systems in the BPASS models \citep[derived from ][]{2017ApJS..230...15M}. These disfavour large mass ratios for massive primary stars, based primarily on observations of the Tarantula Nebula, as might be expected if only massive companions survive the pre-MS phase. However the metallicity evolution of these distributions remains uncertain, as does any impact on the pre-main sequence lifetimes of those companions which do exist.  

While simulations of star forming regions continue to improve, alternate approaches  fit photometric data from observed populations including protostars \citep[e.g.][]{2017A&A...604A..78R,2020A&A...636A..54V}. As observational sensitivity in the near-infrared improves, population synthesis including pre-MS objects may allow the impact of massive star feedback on low mass companions to be more robustly evaluated. We note that, at present, we do not include any emission component for our protostars in our spectral synthesis populations. While the BPASS framework allows for both the pre-main sequence lifetime prescription and the emission contribution to be varied in potential future studies, we caution that doing so would be subject to uncertainties in the heavy dust enshrouding of pre-MS stars, and the impact on that dust attenuation of radiation feedback from neighbouring stars.

\section{Conclusions}\label{sec:conclusions}

In this work, we have explored the impact of including pre-main sequence delays between the onset of star formation and the zero age main sequence, in the context of a binary stellar population and spectral synthesis code, and with a focus on impacts in interpreting distant galaxies.

\begin{enumerate}
    \item We implement the mass and metallicity-dependant pre-main sequence lifetimes of \citet{2011A&A...533A.109T} in v2.3 of the Binary Population and Spectral Synthesis (BPASS) framework \citep{2018MNRAS.479...75S,2023MNRAS.521.4995B}. Where binaries are considered, the delay timescale is fixed by the mass of the most massive component.

    \item We find that, as expected, the rest-frame ultraviolet continuum luminosities and ionizing photon production rates of simple stellar populations show very little impact of the stellar pre-main sequence.

    \item Approximately 10 per cent of the rest-frame optical (5000\AA-1$\mu$m) flux density of a 1\,Myr old stellar population originates from stars in the range $1<M/M_\odot<5$. This emission is absent in the pre-main sequence delay scenario.  

    \item The luminosity density at 1 micron (rest-frame) remains dominated by massive stars until stellar population ages $>100$\,Myr, in both the coeval ZAMS and pre-MS delay scenarios and so is a poor proxy for mass at population ages less than this. 

    \item The observed spectrum of distant galaxies is unlikely to be significantly affected by pre-main sequence delays, with the largest impact on the stellar continuum, but care should be exercised when giving results to a high apparent precision.

    \item The mass to light ratio of galaxies varies strongly with population age, and is sensitive to whether stellar mass is defined to include protostars, includes remnants or is limited to stars undergoing nuclear burning. 

    \item The concept of stellar mass, as applied in SED fitting and mass to light ratios is poorly defined, and caution should be exercised when reporting the in-situ stellar mass in the distant Universe. We recommend that stellar mass estimates reported in the literature should specify whether they include proto-stars, whether they refer to the living (i.e. nuclear burning) mass of stars, whether compact remnant masses or included, or whether they refer to the mass of all stars formed since the onset of galaxy formation.

 \end{enumerate}

\section*{Acknowledgements}

We thank Prof. Jan Eldridge for helpful discussions and for her continuing collaboration on the BPASS project. AU is supported by a Warwick Prize PhD Scholarship. ERS and CMB acknowledge funding from the UK Science and Technology Facilities Council (STFC) through Consolidated Grant Number ST/X001121/1.

Figures in section \ref{sec:example} made use of the "ColourBlind" colour table for IDL \citep{2017zndo....840393W}. 

\section*{Data availability}

The data to reproduce each figure, and the underlying population synthesis models, can be made available by reasonable request to the first author.


\bibliographystyle{mnras}
\bibliography{PreMS-BPASS} 

\bsp	
\label{lastpage}

\end{document}